\documentclass[12pt,a4paper]{article}
\pdfoutput=1
\usepackage[utf8]{inputenc}
\usepackage[T2A,T1]{fontenc}
\usepackage[english]{babel}
\usepackage{amssymb,amsfonts,amsmath,mathtext,cite,enumerate,float} %
\usepackage[dvips,pdftex]{graphicx}
\usepackage{geometry} %
\usepackage{ifpdf}
\usepackage{hyperref}  
    \title{Two-dimensional model of intrinsic magnetic flux losses in helical flux compression generators}
    
    \author{V.~V.~Haurylavets\thanks{E-mail:bycel@tut.by}, V.~V.~Tikhomirov\thanks{E-mail:vvtikh@mail.ru}}

\begin{document}
         \maketitle
         \begin{center}
                  Research Institute for Nuclear Problems, Belarusian State University,\\
Bobruiskaya 11, 220030 Minsk, Belarus  
         \end{center}

    \fontsize{14pt}{18pt}
    \selectfont
    \begin{abstract}
    Helical Flux Compression Generators (HFCG) are used for generation of mega-amper current and high magnetic fields. We propose the two dimensional HFCG filament model based on the new description of the stator and armature contact point. The model developed enables one to quantitatively describe the intrinsic magnetic flux losses and predict the results of experiments with various types of  HFCGs. We present the effective resistance calculations based on the non-linear magnetic diffusion effect describing HFCG performance under the strong conductor heating by currents. 
    \end{abstract}
    \section{Introduction}

Helical flux compression generators (HFCGs) are compact pulsed
 power sources of current and voltage. The interest in HFCG studies primarily
 stems from their unique capability to achieve very high energy
 densities,  magnetic field strengths  and to generate high-power current pulses.
They find promising applications  in particle accelerator
technology, magnetic plasma confinement,  neutron radiation pulse
and  Z-pinch generation, as well as in thermonuclear research. In
this field, they are a relatively cheap option to large stationary
mega-ampere  pulsed current generators.  There is an interesting
proposal to use HFCGs with nuclear explosives for developing
accelerators having a short operating time and a very large beam
luminosity at the energies only recently available in modern
accelerators. Moreover, based on HFCG, it is possible to create a
large pulsed magnetic lens for focusing proton beams of intensity
$10^{23}$ protons per second per surface area of about 1~mm$^2$
\cite{Sakharov}.

The idea of  the FCG was first proposed and substantiated in 1951
by A. D. Sakharov \cite{Sakharov}, who suggested converting the
chemical energy of explosives into  magnetic field energy. His
suggestions were implemented in 1952 in the MK-1 experiments on
magnetic flux compression performed in VNIIEF
 (Russian Federal Nuclear Center  All-Russia Research Institute of Experimental
 Physics). In 1952, M. Fowler in the United States compressed the magnetic field
 using  his first plate generator \cite{Fowler}. Many countries have
  joined in the research on FCGs ever since.

  Some difficulties were encountered in theoretical description of the
  HFCG. Theoretically predicted currents were several times as large as those
    measured  experimentally   \cite{Kiuttu}.

   Various empirical factors affecting the magnetic flux losses were used
   to eliminate the discrepancies between  theoretical
  estimates and experimental data. As a rule, these factors were not
  of universal character, and thus applied to certain HFCG designs or operation
parameters. For example, the resistance could be
  increased by a factor of two and more, depending on the HFCG design, the
  load used, and the initial current.
    A number of HFCG numerical
  models using these  factors have been proposed over the years.
  The awareness of the fact that besides the resistance
  losses, there exist other
  losses in the vicinity
  of the contact point appeared to be the only consistent view
  \cite{Neuber, Kiuttu, Crawford}.

  The numerical model  suggested in the present paper enables
  describing the parameters of operating HFCGs.
  It is shown that because of the lack of comprehension about the nature and effects of
  magnetic flux losses, some of them were neglected earlier. The developed numerical  model
  of the HFCG moving contact
  point  gives the insight into the nature of these losses and allows
  computing  their values.

\section{Geometry and Operating Principle}
    { HFCG is a device compressing a magnetic flux.
    The magnetic flux $\Phi$ passing through the surface $S$ can be found using the
    surface integral:
            \begin{equation}
\Phi=\int_S\textbf{B}\cdot d\textbf{s}, \label{eq:magnetic_flux}
\end{equation}
    where $B$ is the magnetic field induction, and if the closed circuit $c$ belongs to the open surface
    $S$, then \cite{Paul}:

                \begin{equation}
\dfrac{d\Phi}{dt}=-\dfrac{d}{dt}\int_S\textbf{B}\cdot ds=\oint_c \textbf{E}\cdot d\textbf{l}, \label{eq:magnetic_flux}
\end{equation}
  where $ \textbf{E}$ is the electric field strength along the closed circuit $c$, $\textbf{E}\cdot
  d\textbf{l}$ shows that here the differential length $
  d\textbf{l}$  the electric field tangential with
  respect to the circuit are used. One can see from
  \eqref{eq:magnetic_flux} that with the conserved magnetic flux
  and decreasing length of the conducting closed circuit $c$, both
  the electric field and the current increase in the circuit. By definition, the
  inductance of the closed circuit having a current $I$ is
                  \begin{equation}
L=\dfrac{\Phi}{I}, \label{eq:inductance_flux}
\end{equation}
        The   physical principle governing the operation of the HFCG is based on
    Faraday's law.
        \begin{equation}
\dfrac{d\Phi}{dt}=-RI, \label{eq:pharadei}
\end{equation}
    where $\Phi$ is the magnetic flux in the HFCG, $R$ is the resistance, and $I$ is the current.
        \begin{figure}[h]
        \centering
        \includegraphics[width=0.7\linewidth]{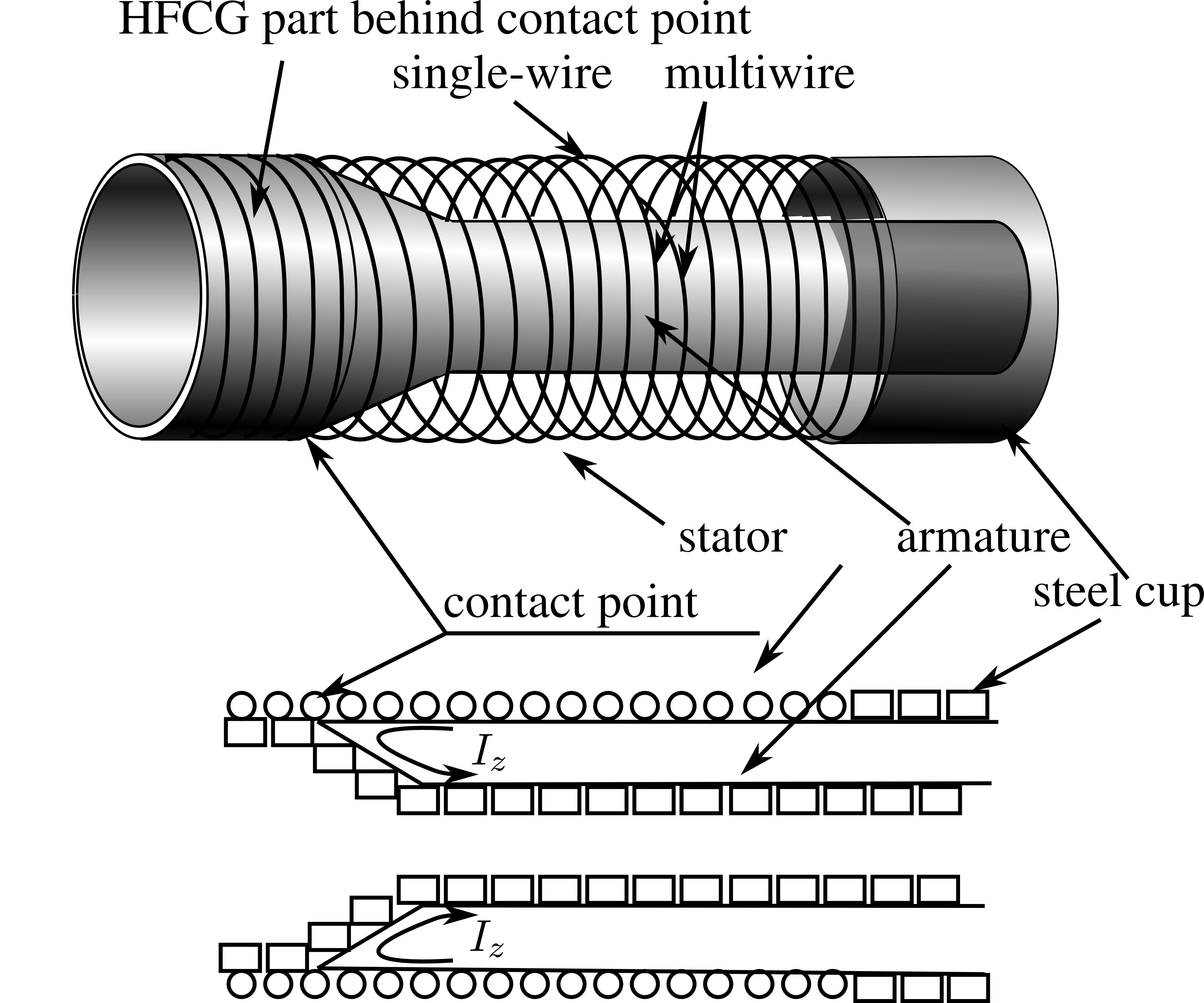}
        \caption{Main parts of the HFCG. Sectional view}
        \label{fig:chema_mkg}
    \end{figure}

    The HFCG design is illustrated in Fig.  \ref{fig:chema_mkg}; its
    main parts are  the  solenoid, called the stator,  and the metal
    tube, called the armature. The stator
     with a single-wire coil  is called the single-wire
     stator; a plural number of
     metal wires helically wound together and arranged in
     parallel relation with each other form a multi-wire stator;
     the number of winds corresponds to the number of wires
     wound in parallel.
    The armature, made of copper or aluminium, is placed inside the stator and is
    filled with explosives. Cups, conducting tubes having the same
    radius as the radius of the stator, are often used in HFCG design: placed at the head, they reduce the effect of
    the armature and stator closure on the HFCG performance, placed at the back, they provide
   the current output to the load.
    The armature and the stator are switched into the circuit and
    connected  through the load, as shown in Fig.
     \ref{fig:chema_raboti_MKG}.

        \begin{figure}[h]
        \centering
        \includegraphics[width=0.5\linewidth]{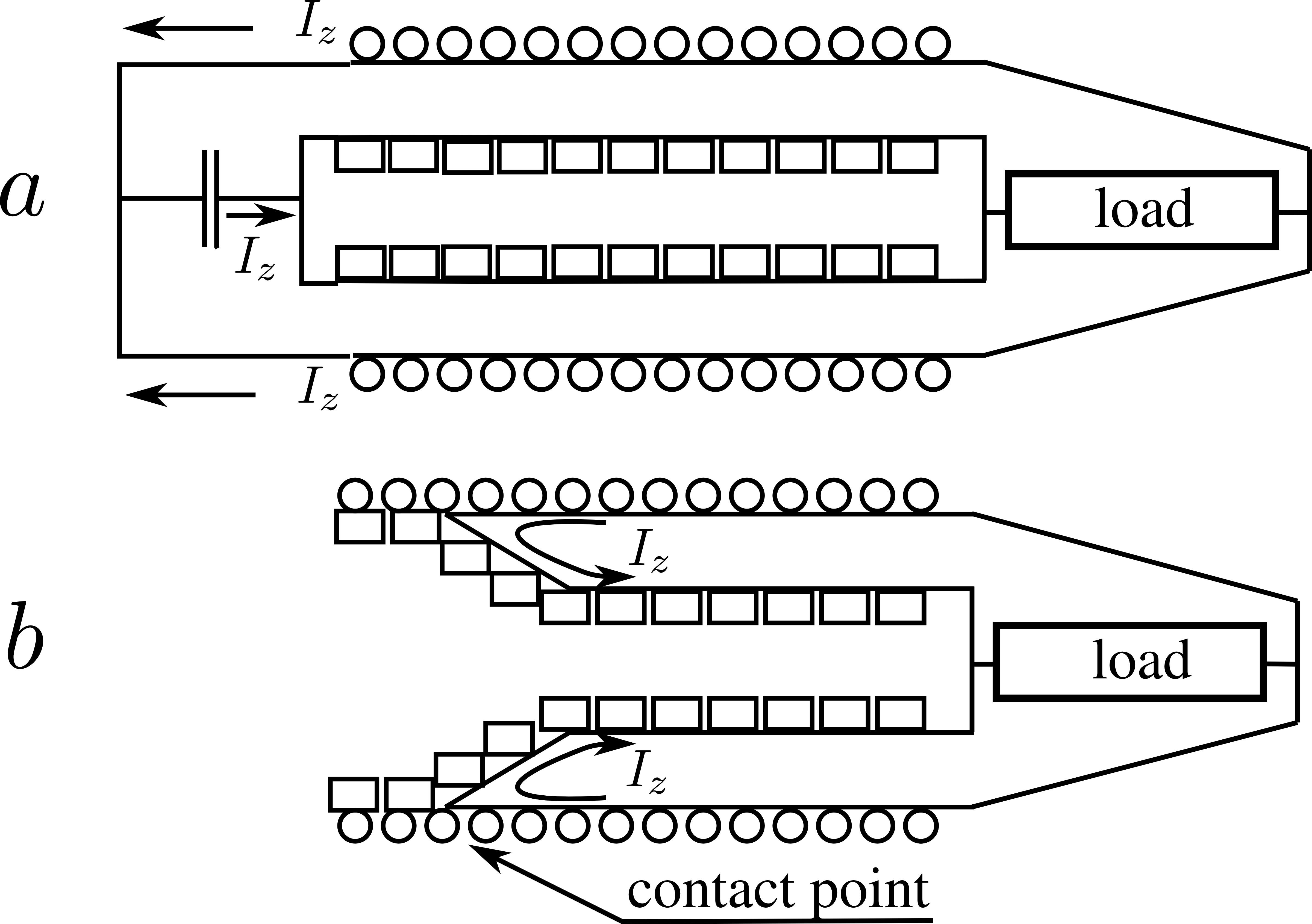}
        \caption{Stages of HFCG operation: \textit{a} is establish initial current,
       \textit{ b} is magnetic flux compression}
        \label{fig:chema_raboti_MKG}
    \end{figure}

The external magnetic field or the initial current in the stator
are used to produce a magnetic flux. The explosives are ignited
from the head of the armature, forcing it to expand. The expanding
armature, forming a conical  plane moving along the length of the
generator, reaches the stator and sequentially shorts out the
turns of the stator.
    }

    \section{Review of the Methods for HFCG Description}
    {
        Let us consider the main approaches to the description of HFCG
        operation.
  In view of (\ref{eq:pharadei}), we have
              \begin{equation}
I\dfrac{dL}{dt}+L\dfrac{dI}{dt}+RI=0. \label{eq:potoc}
\end{equation}
 The solution to  (\ref{eq:potoc}) is
            \begin{equation}
I=I_0\dfrac{L_0}{L}e^{-\int\frac{R}{L}dt}.
\end{equation}

All electrotechnical models, as a rule, can be reduced to equation
(\ref{eq:potoc}), though the methods used to  calculate the
inductance or the resistance may vary. But the models based on the
solution of the equations of magnetohydrodynamics are
fundamentally different. The models based on (\ref{eq:potoc}) are
called zero-dimensional (0D) when the inductance and resistance
are defined by certain functions of time, and one-dimensional (1D) when
the inductance and  the resistance are certain functions of the
coordinates along the HFCG axis and  are calculated at every
integration step.

A two-dimensional HFCG model was suggested in \cite{Novac} and
described in \cite{Fortov}.
   The current-carrying elements of the HFCG are decomposed
   as shown in Fig.  \ref{fig:chema-uprosh-analiza} (see \cite{Novac}).
 The helical
   FCG is decomposed into equivalent
    $z$- and $\theta$
    current-carrying circuits.
    The stator-armature-load electric circuit consists of a coaxial
    part and z-circuits connected to the load.

     The stator is
    decomposed into $N$ number of rings equal to the number of the solenoid
    turns, through which the load current flows in the azimuthal
    direction, inducing the  axial magnetic field $B_{z}^{z}$; the coaxial part of the generator produces
    the field $B_{\theta}$. To correctly describe the $\theta$-current in the
    armature, which is induced  by a changing current $I_{z}$ in
    the stator, let us consider several separate
    $\theta$ circuits taken as separate rings equal in number to
    the number of the solenoid turns, $N$, as  shown in Fig.
    \ref{fig:chema-uprosh-analiza}, and arranged in the
    equivalent electric circuit as  shown in Fig.
    \ref{fig:Ecvivalent-circle}). The current $I^{\theta}_i$ induces the axial magnetic field $B^{\theta}_z$.

    \begin{figure}[h]
        \centering
        \includegraphics{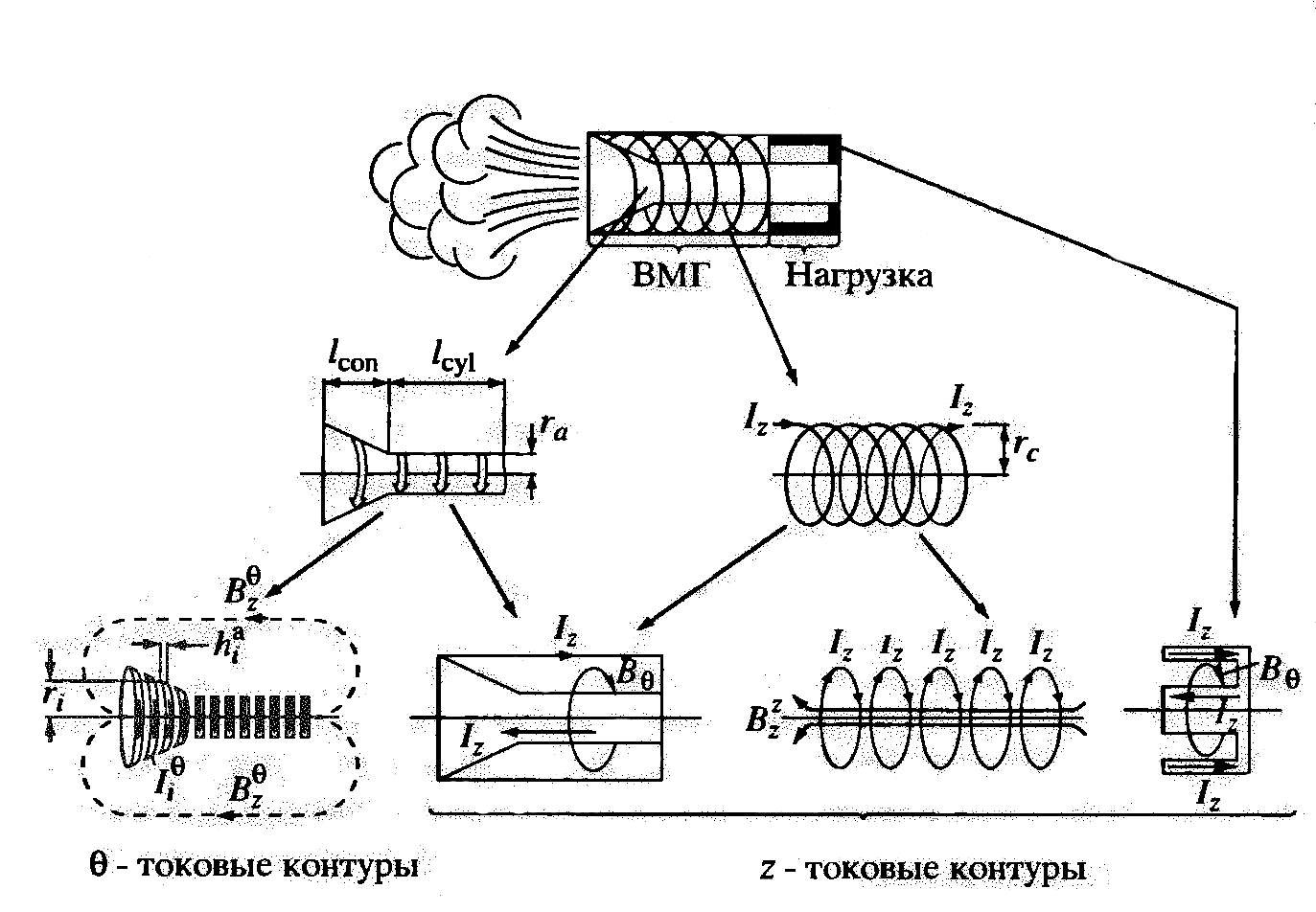}
        \caption{Schematic diagram of HFCG analysis.
        Decomposition of the stator and the armature into  $\theta$ and  $z$ current-carrying circuits}
        \label{fig:chema-uprosh-analiza}
    \end{figure}

    \begin{figure}[h]
        \centering
        \includegraphics{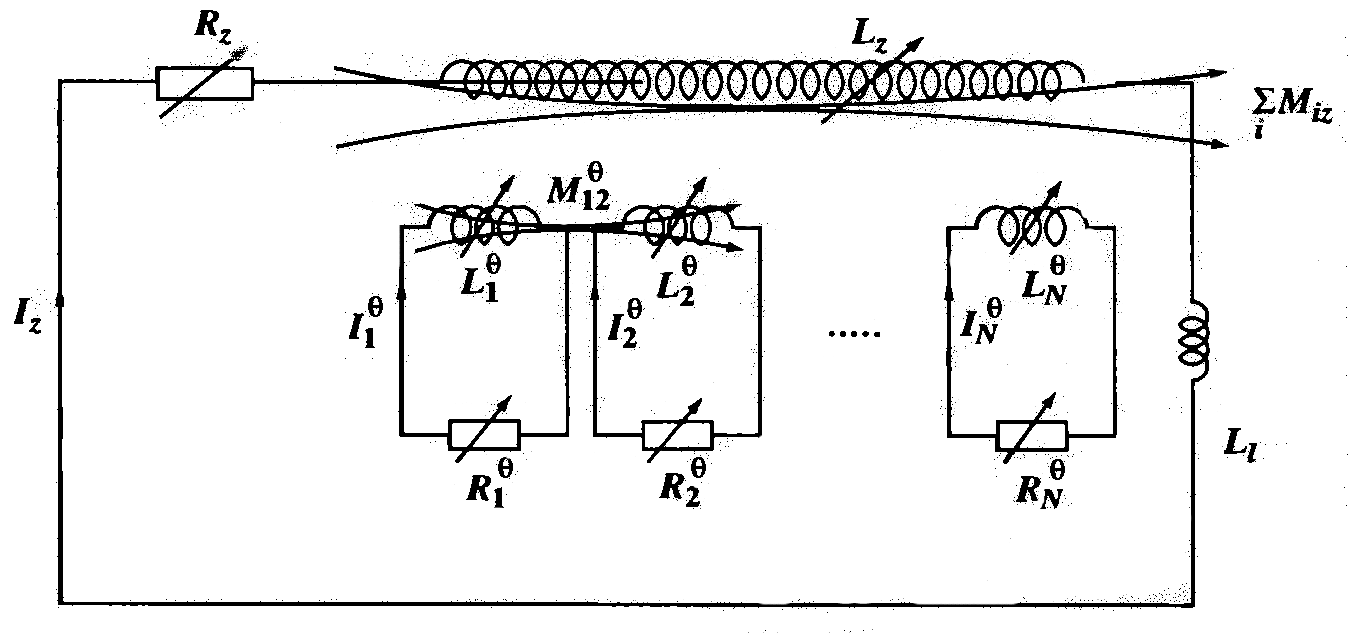}
        \caption{Equivalent electric circuit for $\theta$ and  $z$ current-carrying circuits.}
        \label{fig:Ecvivalent-circle}
    \end{figure}

    The system of equations given in \cite{Novac} for the decomposed circuits has the form (\ref{eq:Iz circuit}) and
    (\ref{eq:theta circuit}).

For the load circuit

    \begin{equation}
    L_{z}\frac{dI_{z}}{dt}+\sum_{i=1}^{N}\left(M_{iz}\frac{dI_{i}^{\theta}}{dt}+I_{i}^{\theta}\frac{dM_{zi}}{dt}\right)
    +\left(R_{z}+\frac{dL_{z}}{dt}\right)I_{z}=0 \label{eq:Iz circuit},
    \end{equation}
    where $L_z$ is the stator inductance, $M_{iz}$ is the mutual inductance between the  $i$-th ring of the armature
    and the stator, and  $I_{i}^{\theta}$ is the current in the $i$-th ring of the armature.
        \begin{equation}
R_z=R^{l}_z+R^z_z+R^p_z+R_{load},
\end{equation}
where $R^{l}_z$ is the resistance of the armature to current
$I_z$, $R^z_z$ is the resistance of the stator, $R^p_z$ is the
resistance describing the proximity effect, which is allowed for
when the diameter of the stator wire is less than the coil pitch
because of the insulation, and  $R_{load}$ is the resistance of
the load.

For $\theta$-circuits, we write $N$ number of equations:
    \begin{equation}
    L_{i}^{\theta}\frac{dI_{i}^{\theta}}{dt}+M_{iz}\frac{dI_{z}}{dt}+I_{z}\frac{dM_{iz}}{dt}+\sum_{j=1(j\neq i)}^{N}
    \left(M_{ij}^{\theta}\frac{dI_{j}^{\theta}}{dt}+I_{j}^{\theta}\frac{dM_{ij}^{\theta}}{dt}\right)+
    \left(R_{i}^{\theta}+\frac{dL_{i}^{\theta}}{dt}\right)I_{i}^{\theta}=0,\label{eq:theta circuit}
    \end{equation}
where $i=1,2,3,...$, $N$ is the number of $\theta$-circuits,
$L_{i}^{\theta}$ is the inductance of the $i$-th ring of the
armature, $M_{ij}^{\theta}$ is the mutual inductance of the
armature rings, and  $R_{i}^{\theta}$ is the resistance of the
$i$-th ring of the armature.

For a better understanding of equations  (\ref{eq:Iz circuit}) and
(\ref{eq:theta circuit}), let us represent them in a compact form:
    \begin{equation}
    \frac{d}{dt}\left(L_{z}I_{z}+\sum_{j=1}^{N}M^{\theta }_{j,z}I^{\theta }_{j}\right)=-R_{z}I_{z}
    \end{equation}
    and
    \begin{equation}
    \frac{d}{dt}\left(M^{\theta }_{i,z}I_{z}+\sum_{j=1}^{N}M^{\theta }_{i,j}I^{\theta }_{j}\right)
    =-R^{\theta }_{i}I^{\theta }_{i},\,\,\,\, i=1...N,
    \end{equation}
    where $M^{\theta }_{i,i}=L^{\theta }_{i}$ is the self-inductance of the $i$-th ring
    if  $j=i$.

B.M. Novac et al.  \cite{Novac} have considered the problem of
the stator turns switching off after the closure of the armature.
In this model, the number of stator turns is the same  as the
number of the armature rings. When the stator turn is closed, the
opposite-lying armature ring is eliminated from the system of
equations  (\ref{eq:Iz circuit}) and  (\ref{eq:theta circuit}).
 The currents in the circuits are recomputed so that the  magnetic fluxes
 before and after the elimination of the turn and the armature ring will  be equal.

An interesting model for magnetic flux loss simulation was
developed by Kiuttu and co-workers in \cite{Kiuttu,KiuttuLast},
where the magnetic field diffusion into the conductor was
considered. By approximating the nonlinear magnetic diffusion and
comparing it with the flux compression rate  one can identify some
distinct regions in the vicinity of the moving contact point,
which are separated by the critical and transition points. At the
critical point, the rate at which the magnetic flux is pushed
ahead by the expanding armature is almost equal to the rate of
flux diffusion into the conductor, and so the magnetic flux after
the critical point is  wasted for compression. Transition point is
the point behind which the stator turn-armature proximity effects
are more important than the turn-to-turn effects.  These effects
lead to arising effective resistance as described by Kuittu. This
approach seems promising because it allows taking account of the
magnetic flux loss at the contact point for magnetohydrodynamical,
0D, and 1D  models. The resistance at the contact point, as
defined by Kuittu, is derived from theoretical considerations and
enables one to describe specific HFCGs, but it does not agree with
the conclusions and computations made in \cite{Neuber}.

The idea suggested and substantiated in \cite{Neuber} is that the magnetic flux losses are concentrated in
the vicinity of the moving contact point and are not related to resistance, as only taking this approach,
one can simultaneously match both the output voltage and the current  of the HFCG with its parameters. The authors of \cite{Neuber}
propose to describe the magnetic flux loss, called the intrinsic loss, using the following equation for the HFCG:
    \begin{equation}
I \cdot \alpha \cdot \dfrac{dL}{dt}+L\dfrac{dI}{dt}+IR=0,
    \end{equation}

  where  $\alpha$ is the flux loss parameter, varying within the interval from 0 to 1.
  It is also stated that though the flux losses cannot currently be computed numerically,
  one can approximate them by the flux loss parameter  $\alpha$ and that  the losses occur
  only after the  appearance of the moving contact point between the stator and the armature.
  In the present
  paper, a different approach is applied to the consideration of magnetic flux losses.  Using this approach,
  one can with satisfactory accuracy describe HFCGs of completely different designs and  with various stators
  on the basis of the physical theory and  pre-experimental data alone, and thus predict the results of the experiments.

    \section{Basic Formulas. Governing Equations}
    {
 In the present paper, a two-dimensional model of the HFCG, based on the model described in \cite{Novac}, is developed.
For correct computation of the current distribution in the
armature, we used the following decomposition: the stator is
decomposed into the turns and the armature is decomposed into the
rings, which are assigned the equivalent current-carrying circuits
(see Fig.\ref{fig:razbivka_MKG}).
     \begin{figure}[h]
        \centering
        \includegraphics{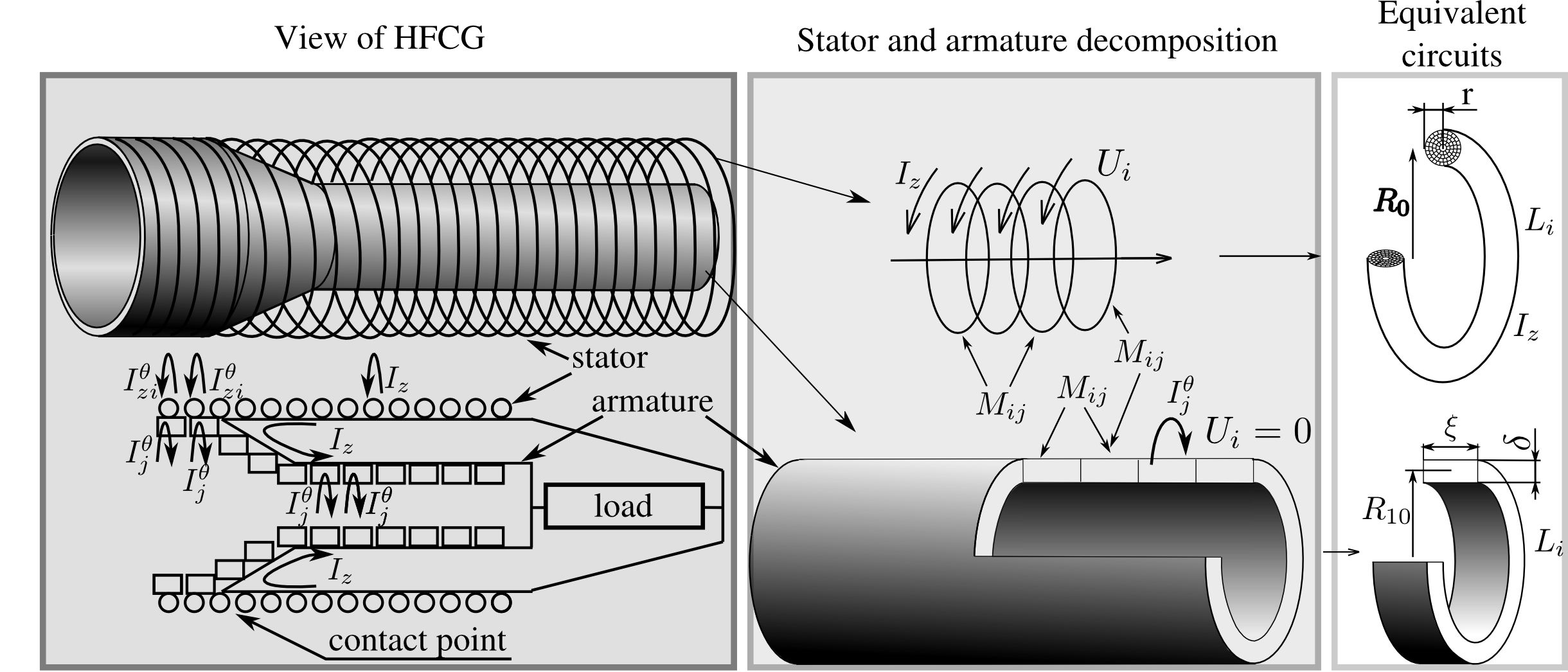}
        \caption{Decomposition of the HFCG elements into the equivalent current-carrying circuits,  $I_z$ is the
        current in the stator-armature-load electric circuit,  $I^{\theta}_j$ is the current in the armature rings,
        $I^{\theta}_{zi}$  are the $\theta$-currents in the stator turns after it has contacted the armature and
        has been
        eliminated from the stator-armature-load electric circuit}

        \label{fig:razbivka_MKG}
    \end{figure}
    Each equivalent circuit is assigned the self-inductance $L_i=M_{ii}$ and the resistance $R_i$. In this case,
    the system of equations for equivalent circuits takes the form:
    \begin{equation}
    \sum_j^nM_{ij}\frac{dI_j}{dt}+\sum_j^n\frac{dM_{ij}}{dt}I_j+R_i I_i=U_i, \label{eq:uravnenie-s-razbivkoj}
    \end{equation}
    where  $U_i$ is the voltage produced in the circuit as a part of the electric circuit (galvanic coupling between
    the turns of the stator), and $M_{ij}$ is the mutual inductance between the equivalent circuits $i$  and  $j$; here  $i,j=1...n$,
    where $n$ is the total number of the current-carrying circuits.
    In solving the system of equations \eqref{eq:uravnenie-s-razbivkoj}, we assume that the rings of the armature
    and the metal cup are not connected with zero voltage across them. The voltage across the load equals the total voltage of the stator turns
        \begin{equation}
    \sum^N_{i=k+1} U_i =U_{load},
    \end{equation}
    where $k$ is the number  of the stator turns switched out of  the electric circuit and at the initial time,
    $k$ equals  zero.  The voltage $U_i$ across the turns switched out of the electric circuits is zero.

    To solve the system of equations  \eqref{eq:uravnenie-s-razbivkoj}, let us recast it in the form:

    \begin{equation}
        \begin{array}{c}
            M_{11}\frac{dI_1}{dt}+M_{12}\frac{dI_2}{dt}+...=-\frac{dM_{11}}{dt}I_1-\frac{dM_{12}}{dt}I_2-...-R_1 I_1+U_1\\
            M_{21}\frac{dI_1}{dt}+M_{22}\frac{dI_2}{dt}+...=-\frac{dM_{21}}{dt}I_1-\frac{dM_{22}}{dt}I_2-...-R_2 I_2+U_2.\\
            \dotfill
        \end{array}
    \end{equation}
    The dependent variables are canceled out; in the elements outside the electric circuit, the unknown quantities $U_i$ are either
    switched out or equated to zero:
        \begin{equation}
        \begin{array}{c}
            L_{z}\frac{dI_z}{dt}+M_{z,N+1}\frac{dI_{N+1}}{dt}+...=-\sum_{i=1}^N\frac{dM_{i1}}{dt}I_1-\sum_{i=1}^N\frac{dM_{i2}}{dt}I_2-...-R_z I_z\\
            M_{z,N+1}\frac{dI_z}{dt}+M_{N+1,N+1}\frac{dI_N+1}{dt}+...=-\frac{dM_{N+1,1}}{dt}I_1-\frac{dM_{N+1,2}}{dt}I_{}-...-R_{N+1} I_{N+1},\\
            \dotfill\label{sistema_uravnenij}
        \end{array}
    \end{equation}
    where $L_z=\sum_{i=1}^NM_{ii}, M_{zi}=\sum_{j=1}^NM_{ji},$ and $ R_z=\sum_{i=1}^NR_i$ are the quantities characterizing the stator included into the
    stator-armature-load electric circuit, $L_z$ is the stator inductance, $M_{iz}$ is the mutual inductance between the  $i$-th circuit  of the armature ring and the stator, and  $R^z_z$ is the resistance of the stator. We thus obtain the system of rank  $n-N+k+1$, where $N$ is the
    number of the stator turns;  for $k=0$ (all the stator turns are included into the electric circuit), the system has rank $n-N+1$,
    with $n$ being the number of  HFCG current-carrying circuits. When the first stator turn is switched out of  the electric circuit, and $k=1$:
        \begin{equation}
        \begin{array}{c}
       M_{1,1}\frac{dI_1}{dt}+ M_{z,1}\frac{dI_z}{dt}+M_{1,N+1}\frac{dI_N+1}{dt}+...=-\frac{dM_{1,1}}{dt}I_1-\frac{dM_{1,2}}{dt}I_{}-...-R_{1} I_{1}\\
           M_{2,1}\frac{dI_1}{dt}+  L_{z}\frac{dI_z}{dt}+M_{z,N+1}\frac{dI_{N+1}}{dt}+...=-\sum_{i=2}^N\frac{dM_{i1}}{dt}I_1-...-R_z I_z\\
            M_{N+1,1}\frac{dI_1}{dt}+M_{z,N+1}\frac{dI_z}{dt}+M_{N+1,N+1}\frac{dI_N+1}{dt}+...=-\frac{dM_{N+1,1}}{dt}I_1-...-R_{N+1} I_{N+1}.\\
            \dotfill
        \end{array}
    \end{equation}

To take account of the load in the electric circuit and the
coaxial part of the HFCG, let us add  the inductance of the load
($L_{load} = const$) and  the inductance of both the armature and
the metal cup  ($L_{line}^z$) to the  inductance of the stator; we
shall also add  the resistance of the load ($R_{load}$) and the
armature and the metal cup ($R_{line}$) to current $I_z$ to that
of  the stator. Then the system of equations takes the form:
    \begin{equation}
        \begin{array}{c}
            (L_{z}+L_{load}+L_{line}^z)\frac{dI_z}{dt}+M_{z,N+1}\frac{dI_N+1}{dt}+...=\\
            =-\frac{dL_{line}^z}{dt}I_z-\sum_{i=1}^N\frac{dM_{i1}}{dt}I_1-\sum_{i=1}^N\frac{dM_{i2}}{dt}I_2-...-(R_z+R_{line}+R_{load}) I_z\\
           M_{z,N+1}\frac{dI_z}{dt}+M_{N+1,N+1}\frac{dI_N+1}{dt}+...=-\frac{dM_{N+1,1}}{dt}I_1-\frac{dM_{N+1,2}}{dt}I_{}-...-R_{N+1} I_{N+1}.\\
            \dotfill\label{reshaemaja_sistema_uravnenij}
        \end{array}
    \end{equation}

   From  \eqref{reshaemaja_sistema_uravnenij} we find the derivative of current for each of the  circuits and
   restore the currents in the circuits using Euler's method:
    \begin{equation}
     I_i=I_i^0+h*\frac{dI_i}{dt},
    \end{equation}
    where  $h$ is the time step.
   The Euler method is applied twice: The resistance is computed
    for each current-carrying circuit and then the current is
    computed for each circuit. The average resistance of each
    circuit during the time step  is computed and then
    the current in each circuit is re-computed.

    The main differences of the model proposed here from that described in   \cite{Novac} are as follows:
    \begin{itemize}
     \item The armature is decomposed into a substantially larger number of rings than the number of rings in the stator;
    \item Account is taken of the arbitrary geometric dimensions of the stator and the  metal cups on the HFCG ends;
    \item The part of the HFCG behind the contact point is modeled instead of being
    ignored;
    \item The stator turn at the contact point is decomposed into smaller parts to achieve  higher accuracy of modeling;
    \item The developed model naturally takes into account the intrinsic flux
    losses;
    \end{itemize}

    It should be noted that for the described
    two-dimensional model, the term  $\frac{dL_{z}}{dt}$ in  \eqref{eq:Iz circuit} can be put equal to zero in the case when in  HFCG modeling
    neither the stator rings are displaced nor their radii
    are changed. In the considered model this occurs spontaneously, since the circuit
    eliminated from the electric  circuit remains there as a free current-carrying circuit. This is basically different from the model
    proposed by Novac and co-authors \cite{Novac}, where the elimination problem leads to a nonzero value of the term $\frac{dL_{z}}{dt}$
    because the current is deliberately eliminated from the closed turns of the stator, while the associated  magnetic flux is artificially re-distributed.

    \section{The Moving Contact Point Model and the Intrinsic Flux Losses in the HFCG}

    {
The region of the moving contact point was modeled in
\cite{Neuber}. The obtained result showed that the current changed
direction after passing the  contact point, flowing from the
stator into the armature. Let us check the result against the
model of the HFCG as a whole.

Let us project the current on the HFCG axis. This gives us the
coaxial part of the HFCG. Let us now project the currents on the
plane perpendicular to the axis. The projections of the currents
coincide with the currents flowing in the equivalent circuits
after the decomposition of the HFCG  conducting elements
(Fig.\ref{fig:razbivka_MKG}).

As is seen in Fig. \ref{fig:razbivka_MKG}, the stator turns are
closed through the armature behind the contact point, and hence
$\theta$-currents may flow through the turns. In terms of magnetic
diffusion, we have two circular rings with adjacent surfaces,
through  which the opposite currents flow. The characteristic time
for the observed  skin depth of $2\cdot10^{-4}$ m is from 1 to 10
microseconds. The same or longer time is required for current
compensation, which is comparable to the HFCG operating time.
Thus, the idea of a rapid current attenuation  behind the contact
point does not describe a real physical picture of HFCG operation.
The idea of slow current attenuation is confirmed by indirect
simulation of  HFCG using (\ref{eq:uravnenie-s-razbivkoj}) and
computing the current density distribution over the HFCG
conductors \cite{We3D}.
        \begin{figure}
    \centering
        \includegraphics[width=0.3\linewidth]{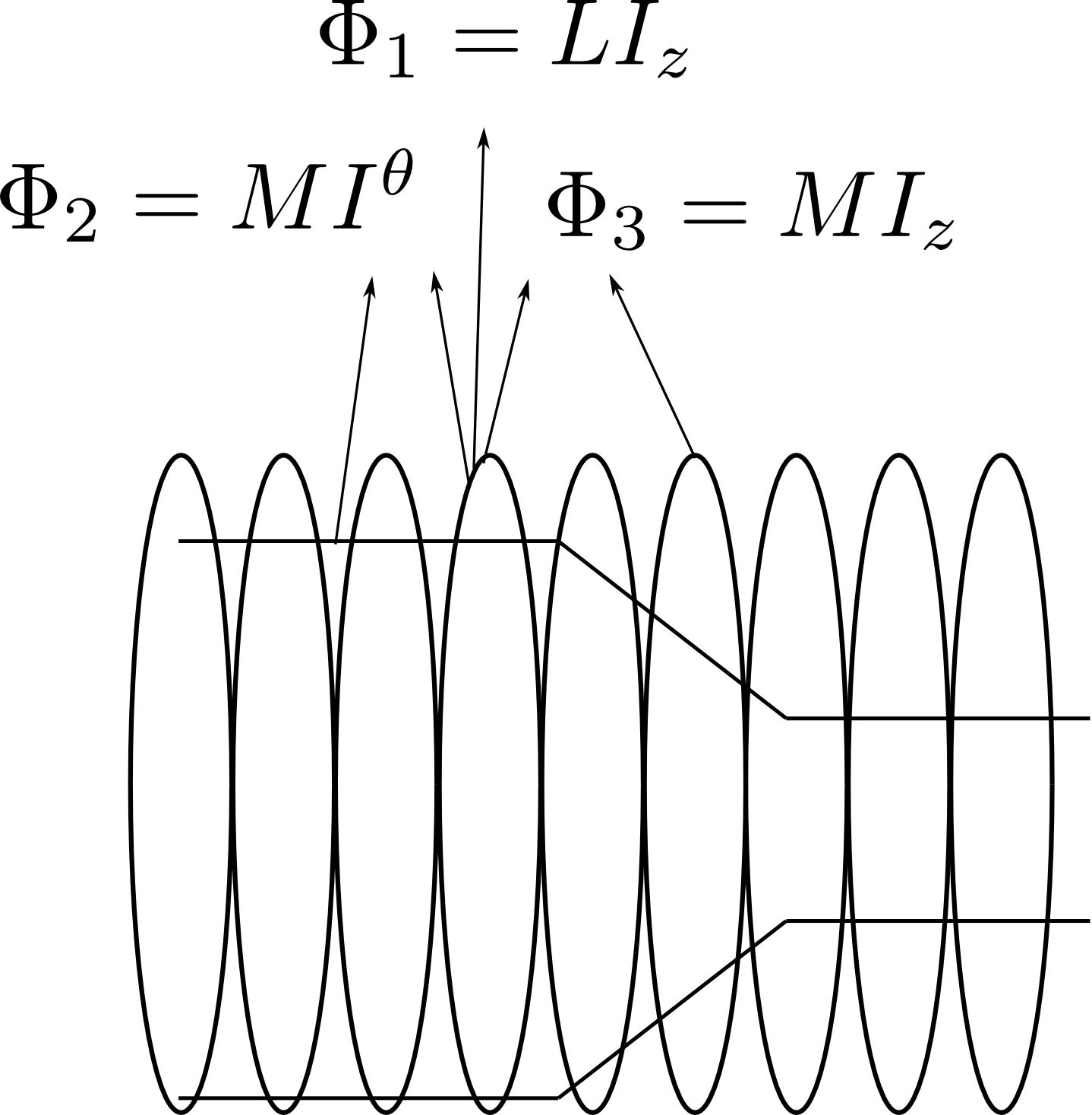}
        \caption{Intrinsic flux loss, the lost magnetic flux}
        \label{fig:I_z}
    \end{figure}
At the moment of the turn closure, the current in the turn is the
same as the current in the stator and is further calculated by the
system of equations (\ref{eq:uravnenie-s-razbivkoj}). The armature
ring stops after it has contacted the stator, and does not move
until the computations are over, so the closed stator turn is
switched out of the stator-armature-load electric circuit. In the
model presented here the closed turn of the stator is considered
as an independent circuit connected to the rest of the circuits
via the magnetic field. The sum of the mutual magnetic flux of the
closed stator turn and the opposite armature rings of the HFCG
behind the contact point is much smaller than the intrinsic
magnetic flux of the stator turn or the armature ring. Hence, the
assumption that the armature ring and the stator turn are fixed
after the closure does not significantly affect the results of
computation.

When the stator turn is closed and switched out of the
stator-armature-load electric circuit, the magnetic flux linked to
it is lost, which includes the flux of the turn itself and the
fluxes of other turns of the stator and of the armature rings.
What remains is the mutual flux between the  switched turn and the
operating part of the stator (Fig. \ref{fig:I_z}). Thus, for
calculating the losses, one needs to consider the flux of the
stator and the armature behind the contact point. For HFCGs with
superconducting generators, this sum flux is equal to zero. In a
real HFCG, the stator turns and the armature rings have different
current densities, as well as the ohmic losses, and so the
magnetic flux in the HFCG behind the contact point is larger the
larger is the difference between the losses in the stator and the
armature. In the model described in \cite{Novac}, the magnetic
flux is artificially redistributed from the closed turns of the
stator into the HFCG part in front of the contact point, leading
to an erroneous representation of the current distribution over
the armature and disregard of the intrinsic flux loss.

    \section{One-Dimensional Model with Regard for the Intrinsic Flux Loss}
    {
   Let us distinguish the  magnetic flux $L_z$ in the
    stator conductor. Recast
    (\ref{eq:potoc}) as
            \begin{equation}
    I\dfrac{dL}{dt}+L\dfrac{dI}{dt}+RI+\dfrac{dL_z}{dt}I=0, \label{eq:pharadei_induct_intrinsic}
    \end{equation}
    which includes the flux taken away  by the turns.
    The solution of (\ref{eq:pharadei_induct_intrinsic}) for HFCG current has the form:
                \begin{equation}
    I=I_0\dfrac{L_0}{L}e^{-\int\frac{R}{L}dt-\int\frac{dL_z}{dt}\frac{1}{L}dt}.
    \end{equation}
    }
\section{Inductance}
    {
For a two-dimensional model, one needs to calculate both the
self-inductances of equivalent circuits and  the mutual
inductances between them.
    Equivalent circuits are presented as circular loops with geometrical mean distance $g_i$ between the area of the wire cross section
     and itself \cite{Kalantarov}. In this case, the inductance of each circuit is defined
    by:

    \begin{equation}
     L=\mu_0 R \left[ \ln\left( \frac{8R}{g}\right) -2\right], \label{eq:inductija_contura}
    \end{equation}
where $\mu_0$ is the permeability of vacuum, $R$ is the radius of
the equivalent circuit (equals the distance between the center of
the circuit and the wire axis), and $g$ is the geometrical mean
distance between the area of the wire cross section and itself.
The geometrical mean distance is
calculated by formula given in \cite{Kalantarov}
\begin{equation}
     ln(g)=\frac{1}{s_1s_2}\iint\limits_{s_1s_2}\ln(\eta)ds_1ds_2,
    \end{equation}

where $\eta$ is the distance between the elementary areas $ds_1$
and $ds_2$. The integration procedure is as follows: each of the
variables $ds_1$ and $ds_2$ is integrated with respect to one
another throughout the entire area and the procedure is repeated.
If supposed that a high-frequency current flows in equivalent
circuits and is uniformly distributed over the surface, then the
geometrical mean distance for the stator turn circuits equals
$g_i=\frac{d}{2}$, $d$ being the diameter of the stator turn wire
\cite{Kalantarov}. For the armature rings and the  metal cup
rings, the logarithm of the geometrical mean distance between the
area cross section and itself equals
    \begin{equation}
     \ln(g)=\ln(\xi+0.0002)+\ln(0.2236),
    \end{equation}

    where $\xi $ is the width of the armature rings and the metal cup rings
    and 0.0002 is the depth $\delta$ of the fixed skin-layer in meters
     (see Fig. \ref{fig:dlj_vzaimnih_induktivnostej}).

      The skin depth can be calculated for the given initial conditions:  the HFCG design, and when heating is important the initial seed
      current is considered. For the solution stability of the system of HFCG equations, the skin depth is considered as a fixed value, and for $\xi $
       considerably larger than the skin depth, the approximation error is negligibly small.
       Account of the variations in the skin depth has little effect on the value of the current derivative and only matters for HFCGs
       with   significantly varying mean operating frequency. Assuming that the symmetry axis of the circuit
        passes through the middle  of the fixed skin depth one can eliminate
        the proximity effects of the stator wires from the consideration and suppose that the  current is uniformly distributed over
         the surface of the stator turns.

    The mutual inductance between two equivalent circuits of the stator turns is calculated as a mutual inductance
     between two infinitely thin circuits through a series representation \cite{Kalantarov}.
     The mutual inductance between the circuits of the rings of the armature  and that between the circuits of the  rings of the metal cup,
     as well as between the circuits of the
     armature ring or the metal cup ring  and the stator turn is calculated from the geometrical mean distance.

    \subsection{Calculation of Mutual Inductance Between Equivalent Circuits from Geometrical Mean Distance}
When the equivalent circuits of the armature and the cup, as well
those of the  armature rings and the stator turns are closely
spaced, it is important that the geometrical dimensions of their
cross sections be considered, which is achieved by taking account
of the geometrical mean distance between the cross sections of the
two circuits.

        The mutual inductance between the equivalent circuits of the rings is calculated as a mutual inductance between two infinitely thin circuits,
    coinciding with the symmetry axes of the cross section of the current-carrying conductor. The distance between the circuits wires
    equals the geometrical mean distance between the cross sections of
    the two circuits(Fig. \ref{fig:dlj_vzaimnih_induktivnostej}
      and Fig. \ref{fig:kolza_sednee_geometri4eskoe}).

    \begin{figure}[h]
        \centering
        \includegraphics[width=0.4\linewidth]{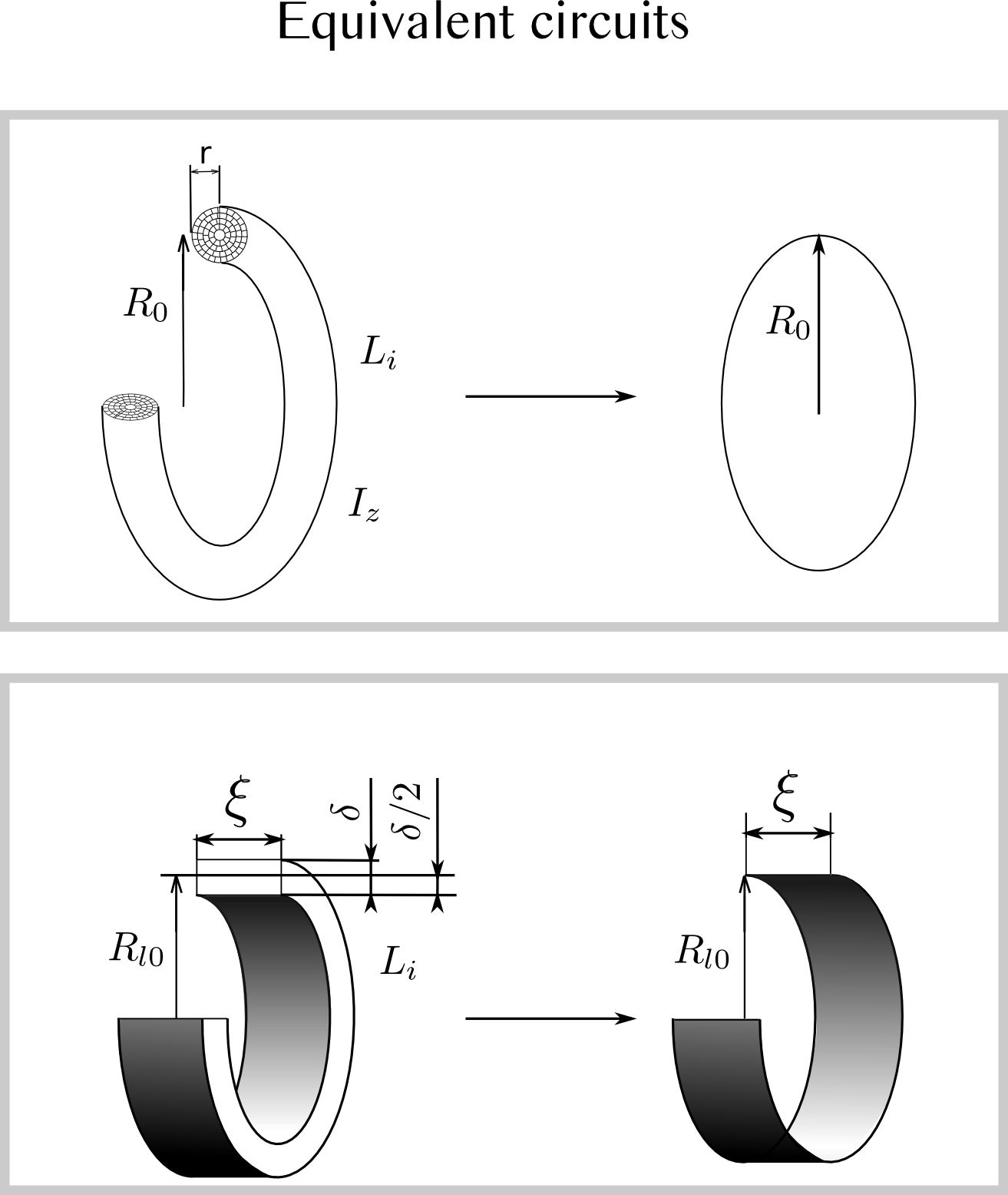}
                \caption{Equivalent circuits for mutual inductance calculation}
        \label{fig:dlj_vzaimnih_induktivnostej}
    \end{figure}

        The circuits of the armature rings and the metal cup rings are represented as the rings of an infinitely thin strip of
        width $\xi$. This approximation is
    correct for skin depths $\delta$ considerably less than the width $\xi$.

    \begin{figure}[h]
        \centering
        \includegraphics[width=0.25\linewidth]{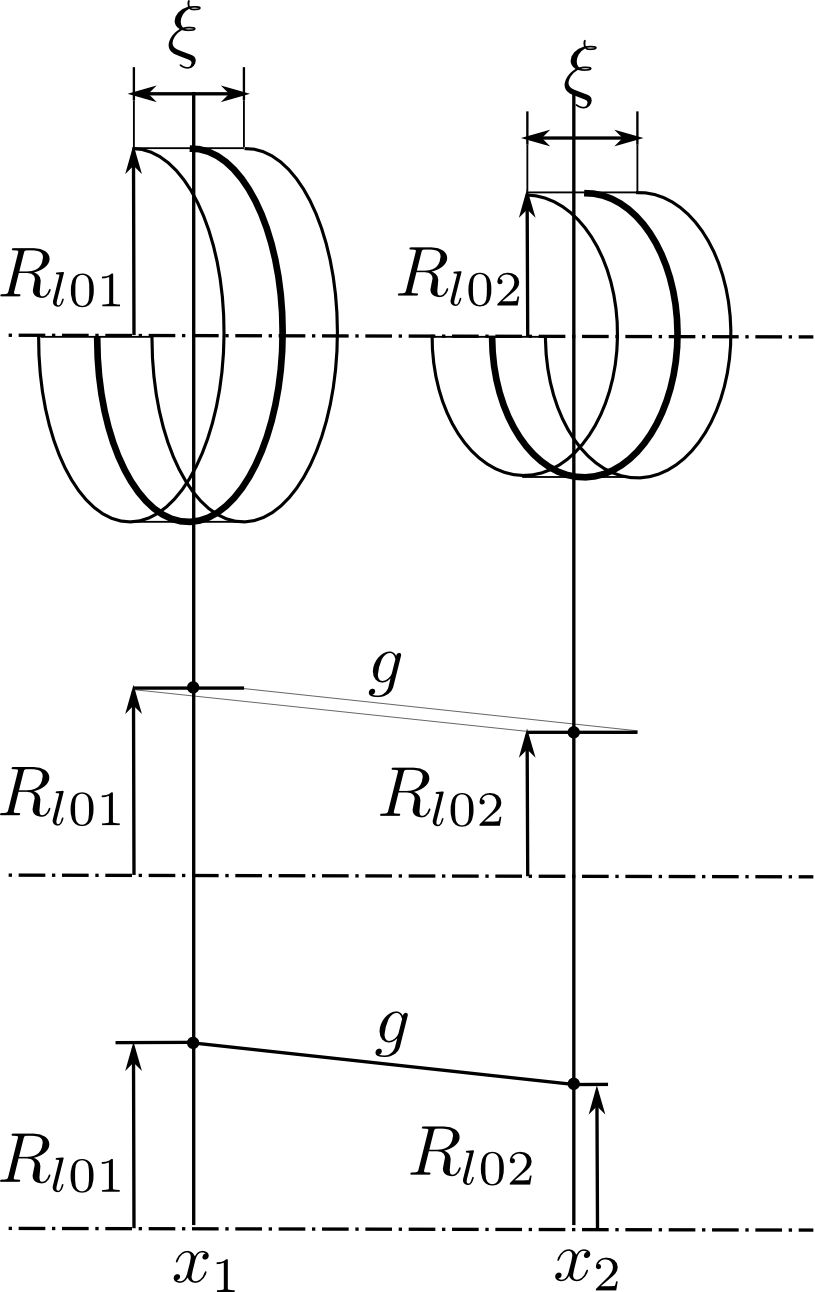}
               \caption{Calculation of mutual inductance between the circuits of the armature
               rings and the metal cup rings from geometrical mean distance}
        \label{fig:kolza_sednee_geometri4eskoe}
    \end{figure}

    Thus, we calculate the geometrical mean distance  between the  cross sections of the armature rings and/or the metal cup rings
        (fig. \ref{fig:kolza_sednee_geometri4eskoe}). The mutual inductance of the circuits is taken equal to  mutual inductance of two infinitely
     thin circuits whose radii are the same as the distance between the center of the ring and the symmetry axis of its cross
     section. The infinitely thin circuits are arranged so that the shortest distance between them is the same
     as the geometrical mean distance between the adjacent cross sections of the
     rings   \cite{Kalantarov}.
     The geometrical mean distance between the cross sections of the armature rings and/or the metal cup rings
     is the geometrical mean distance between two
     segments    (fig. \ref{fig:kolza_sednee_geometri4eskoe}). The geometrical mean distance between the cross sections of the stator turn
     and the armature rings or the metal cup rings is calculated as the geometrical mean distance
      between the point on the symmetry axis of the  cross section of the
      stator turn and
     the segment of the cross section of the armature ring  or the ring of the metal cup.

    \subsection{Decomposition of a Multiwire Stator with Symmetrical Wires into Current-Carrying Elements}
The approach that we suggest is based on the decomposition of a
stator formed by symmetrically arranged wires for the
simplification of the computation procedure and reduction of
errors in calculating the inductance of the  stator.
    Let us make use of this fact that in the stator coil  formed by several wires wound together and
    arranged symmetrically with each other, the same current passes through every wind due to symmetry.

    \begin{figure}[h]
        \centering
        \includegraphics[width=0.4\linewidth]{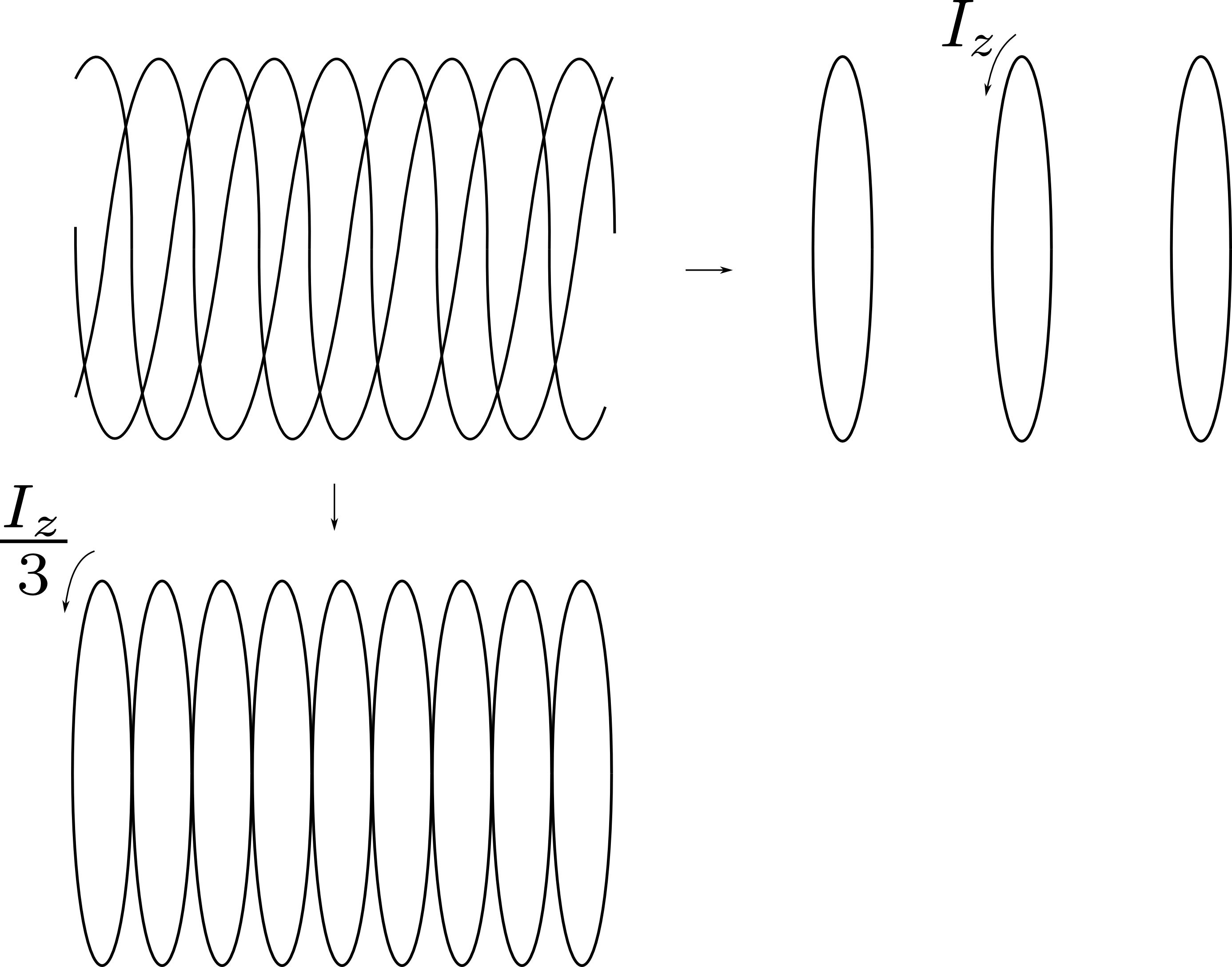}
               \caption{Decomposition of a multiwire stator with symmetrical wires }
        \label{fig:obmotca_zahodi}
    \end{figure}
    The stator turns are decomposed as follows: every  wind has the same number of equivalent circuits as the number of winds.
    The dimensions of the equivalent circuits are assumed to be equal to the dimensions of the stator
    turn (the turn diameter and the wire diameter), as  shown in  Fig.\ref{fig:obmotca_zahodi}. In modeling
    HFCG operation, it should be taken into account in electrotechnical equations that the current passing
    through the obtained circuits is $n$ times smaller than the current $I_z$. When the HFCG consists of the
    sections with different winds and wire diameters, every section is decomposed separately in consecutive order and is considered
    in the system of equations (\ref{eq:uravnenie-s-razbivkoj}).

    \subsection{Decomposition of the Stator Turn into Parts of Equivalent Circuits and Computation of
    their Inductance in the Vicinity of a Moving Contact Point}

Decomposition of the  stator turns into equivalent circuits
provides a helpful tool for describing the HFCG operation. At the
final stage of HFCG operation, switching the circuit of the stator
turn out of the stator-armature-load electric circuit  leads to an
appreciable change in the inductance of the HFCG stator, which
result in a poorer modeling accuracy. For this reason, it is
advisable that a more exact computation of the inductance be
performed at the final stage of HFCG operation, when only several
(ten or fewer) turns remain in the stator.
    \begin{figure}[h]
        \centering
        \includegraphics[width=0.4\linewidth]{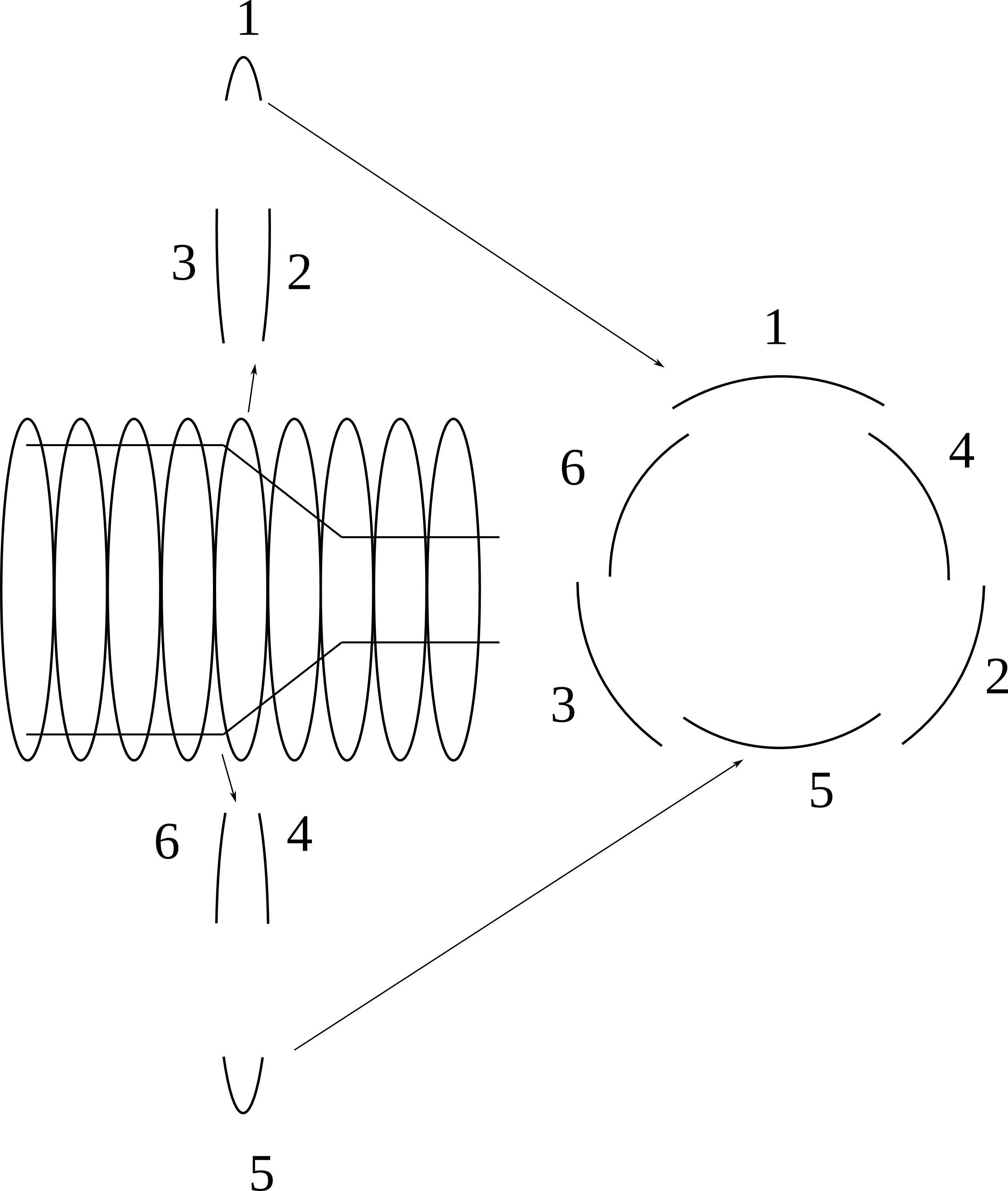}
                \caption{Decomposition of the stator in the vicinity of a moving contact point}
        \label{fig:razbienie_otdelnogo_vitca}
    \end{figure}
    The exact calculation of the stator inductance is performed by
dividing the equivalent circuit of the stator turn into parts
    at the contact point. This is made for both a single wind and
    several symmetric winds. With this aim in view, one can use \cite{Kalantarov}:
    \begin{equation}
    M_{14}=\frac{1}{2}\left( L_{14}-L_1-L_{4}\right),
    \end{equation}
    where $M_{14}$ is the mutual inductance of the two contacting parts of the
    wire that lie on the line curved along the circular arc; $L_{14}$
    is the inductance of the considered segment of the arc  $14$; $L_{1}$ and
    $L_{4}$ and the inductances of its parts; and
    \begin{equation}
    M_{12}=\frac{1}{2}\left( L_{142}+L_{4}-L_{14}-L_{42}\right),
    \end{equation}
    where $M_{12}$ is the mutual inductance of the two non-contacting parts of the
    wire that lie of the line curved along the circular arc;
    $L_{142}$is the inductance of the wire $142$; $L_{4}$
    is the inductance of part $4$; $L_{14}$ is the combined inductance of
    parts $1$ and $4$; and $L_{42}$ is the combined inductance
    of parts $4$  and $2$.
The inductance of the wire composed of three parts is
      \cite{Kalantarov}
    \begin{equation}
    L=L_1+L_{4}+L_2+2\left( M_{14}+M_{42}+M_{21}\right), \label{eq:induct_part}
    \end{equation}
     where $L$ is the inductance of  part  $142$.
     The formula for calculating the self-inductance of the
     wire curved along the circular arc \cite{Kalantarov} reads:
       \begin{equation}
     L=Z-G+A-Q, \label{eq:inductance_of_line}
    \end{equation}
    where $Z$ depends only on the shape and size of the wire axis,  $G,A,Q$  depend on the wire
    cross section and the current distribution pattern over the
    cross section; $G$ also being dependent on the length of
    the wire axis.
    If we neglected the quantities of the order of $\frac{g}{2R_m}$ and
    $\frac{g}{l}$ with $l$ being the wire length and $R_m$, the
    least radius of curvature of the wire axis, then
    \eqref{eq:inductance_of_line} can be written in the form:

    \begin{equation}
     L=Z-G\label{eq:inductance_of_line_pribligenoe}
    \end{equation}
    and
    \begin{equation}
     Z=\frac{\mu_0R}{2\pi}\left[ \theta\left( \ln{8R}-2\right)-4{I'}+4\sin{\frac{\theta}{2}} \right],
    \end{equation}
    where $\theta$ is the  angle subtending the arc of length equal to the total length of the wire,
    while
    \begin{equation}
    {I'}=-\int\limits_{0}^\frac{\theta}{4}\ln{\tan{\vartheta_1}d\vartheta_1},
    \end{equation}
    and
    \begin{equation}
     G=\frac{\mu_0l}{2\pi}\ln{g}.
    \end{equation}
    Thus, using the contact point, it is possible to divide the equivalent circuit of the
    turn into parts and define their self- and mutual inductances.

The contact point divides the equivalent circuit of the turn of a
single-wire stator into two parts, one of which is included into
the common stator-armature-load electric circuit, while the other
one is switched out of  the circuit and is considered as a free
circuit interacting with other circuits via the magnetic field
(Fig. \ref{fig:razbienie_otdelnogo_vitca}). Using the above
formulas, let us calculate self-inductances of the parts of the
turn and mutual inductances between them and other circuits.

By way of example, let us consider a three-wire stator; similar
 calculations for the case of a multiwire stator can be easily
made by generalizing the above theorems and formulas. Because for
symmetric winds of the stator we used the above-described method
of division into the turns, the contact point divides the
equivalent turns into number $2n$ parts (Fig.
\ref{fig:razbienie_otdelnogo_vitca}).

Thus, using the above approach, one can define the self-inductance
of a part of the stator turn bounded by a moving contact point
(Fig. \ref{fig:razbienie_otdelnogo_vitca})
 in the stator-armature-load electric circuit as follows:
     \begin{equation}
     L=3L_{1}+6M_{12},
    \end{equation}

    outside the circuit
        \begin{equation}
     L=3L_{4}+6M_{45}.
    \end{equation}

    The mutual inductance between the parts equals:
    \begin{equation}
     M=3M_{15}+6M_{14}.
    \end{equation}

    \subsection{Inductance of the Coaxial Part of the HFCG}

    It is important that in HFCG operation,  the coaxial inductance of the  stator that is associated with  current
    $I_z$  should be considered.

    Though this inductance is small, it is vital to consider it at
    the final operation stage, when a few turns are left in the
    circuit, and small inductance values, which can be neglected at the initial stage,  play an important
    role.

The projection of the current $I_z$ on the HFCG axis is equal to
itself (see a similar fluid flow problem in hydrodynamics). The
stator is replaced by an equivalent cylinder, with current $I_z$
flowing along its axis (Fig. \ref{fig:truba_dlj_obmotki}), which
is furnished with a metal cup having a diameter approximately
equal to the diameter of the equivalent cylinder.  The cylinder
wall is assumed to be infinitely thin, and the current density is
uniformly distributed over its surface. The radius of the
equivalent cylinder is assumed to be equal to the inner radius of
the stator.

    \begin{figure}[h]
        \centering
        \includegraphics[width=0.7\linewidth]{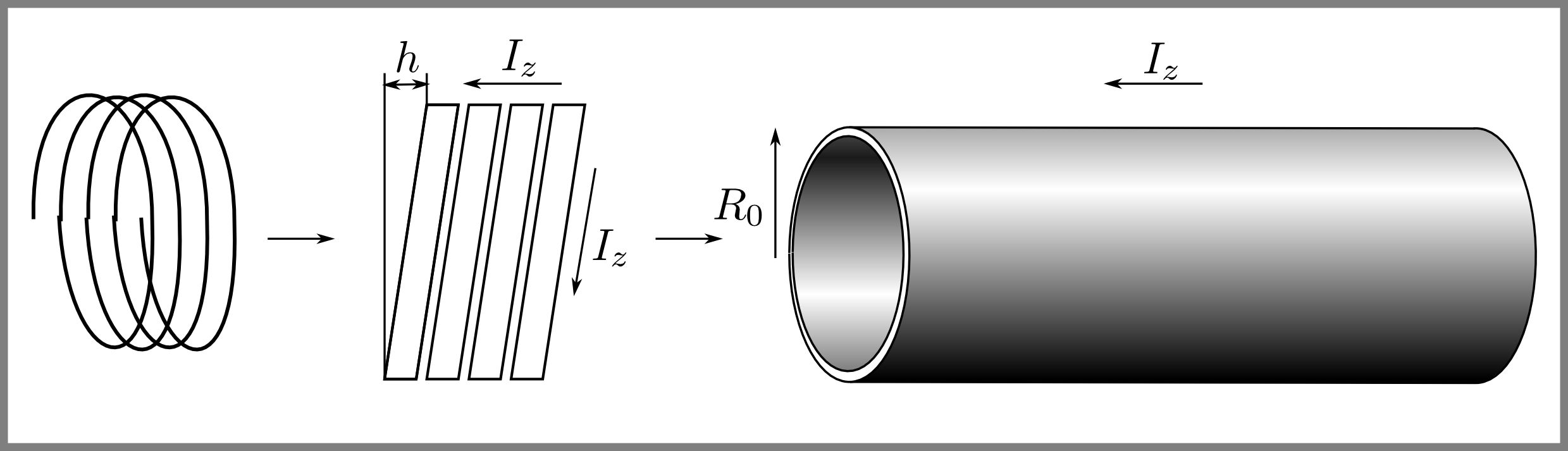}
                \caption{Inductance of the stator for the current projected on the HFCG axis}
        \label{fig:truba_dlj_obmotki}
    \end{figure}

   The self-inductance of the equivalent cylinder can be found in a similar manner as the inductance of
   a straight wire. Here the  geometrical mean distance between the  cross sections of the
   cylinder  and itself equals its radius.

        \begin{equation}
L=\dfrac{\mu_0l}{2\pi}(ln(\dfrac{2l}{r}-1)),
    \end{equation}
    where $r$ and $l$ are the radius and the length of the cylinder, respectively.
    The self-inductance of the armature is determined similarly.

    It is possible to consider a stator with a liner in the form of
    a coaxial cable. In this case, the inductance of
    the HFCG part with a pipe-shaped armature  and that of
    the HFCG part with  a cone-shaped widening armature should be
    calculated separately.

    \begin{equation}
     L_l=L_k+L_m,
    \end{equation}
    where $L_k$ is the inductance of the HFCG with a cone-shaped armature and  $L_m$ is
    the inductance of the HFCG with a pipe-shaped  armature; the HFCG part behind the contact point is  dropped from the consideration.

   The inductance for the current projection on the HFCG axis can be found with known magnetic
   flux:
        \begin{equation}
     L_{line}^z=\dfrac{1}{i^2}\int_s \Phi di,
    \end{equation}
    where $\Phi$ is the magnetic flux, which equals
     \begin{equation}
     \Phi=\int_S B dS.
    \end{equation}
Here $B$ is the magnetic field inductance, $S$ is the area of the
closed current-carrying circuit with current $i$. Then we have
     \begin{equation}
     L_{line}^z=\dfrac{1}{i^2}\int_s \int_S B dS di.
    \end{equation}

When the symmetry of the coaxial part of the HFCG is taken into
account, its inductance in cylindrical coordinates can be found by
formula given in  \cite{Novac}
      \begin{equation}
     L_{line}^z=\dfrac{\mu_0}{2\pi}\int_0^l \int_{rarm(z)}^{rstat(z)} \dfrac{1}{r} dr dz,
    \end{equation}
    where  $rarm(z)$ is the radius of the armature,  $rstat(z)$ is the radius of the stator,  $z$ is the length of the
    coaxial part of the HFCG.

    }
    \section{Errors Related to Neglecting the Stator Helicity}

    When the stator helicity is taken into account, the current $I_z$ can be decomposed into two
    components: along the OZ-axis, which is the generator's axis, and in the XY-plane. The
    projection of the
     current on  the XY-plane equals the current $I_z$ and so does  its projection on the OZ-axis.
        \begin{figure}[h]
        \centering
        \includegraphics[width=0.4\linewidth]{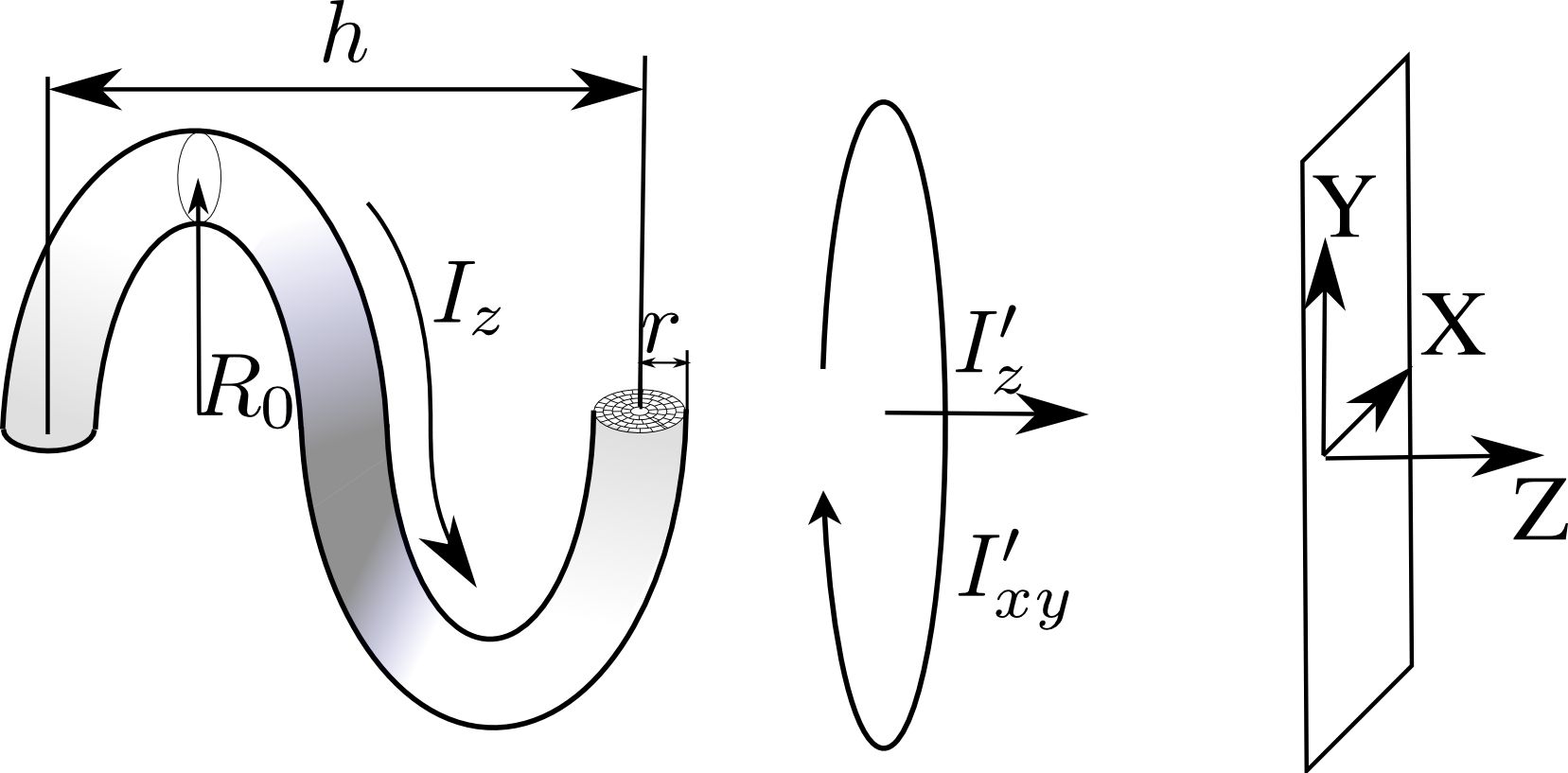}
                \caption{Decomposition of the current $I_z$ into the current azumuthal along the OZ-axis (the HFCG
        axis) and the current in the XY-plane }
        \label{fig:razbienie_tocov}
    \end{figure}

    Neglecting the stator helicity, one can readily find the inductance of the turn in the XY-plane
    by formula  \eqref{eq:inductija_contura}.

    Let us evaluate the helicity effect on the calculations of the stator inductance.
    For the case of uniform distribution of the current density over the wire cross section,
    the formula for calculating the self-inductance can be written as follows:
    \begin{equation}
     L=\frac{1}{s^2}\iint\limits_{s_1 s_2}\overline{M_k}ds_1ds_2,
    \end{equation}
    where
    \begin{equation}
     \overline{M}=\frac{\mu_0}{4\pi}\iint\frac{dl_1dl_2\cos\theta}{D}.
    \end{equation}
    Here it is assumed that $\theta$ is the angle between the length elements $dl_1$ and $dl_2$ in the
    XY-plane. For  rings
    \begin{equation}
     D=2R\sin\frac{\theta}{2},
    \end{equation}
    with due account of the helicity
      \begin{equation}
     D=\sqrt{\left((2R\sin\frac{\theta}{2}\right)^2+\left(\frac{h}{2\pi}\theta\right)^2},
    \end{equation}
    where $h$ is the coil pitch.

    Disregarding the helicity of the
    stator in finding the mutual inductance between the stator turns, we have
    \begin{equation}
     D=\sqrt{\left( 2R\sin\frac{\theta}{2}\right) ^2+\left( nh\right) ^2},
    \end{equation}
    where $n=|i-j|$ and  $nh$ is the distance between the centers of the turns  $i$ and $j$ when  the coil pinch is constant.

    When the helicity is taken into account, we have:
    \begin{equation}
     D=\sqrt{\left( 2R\sin\frac{\theta}{2}\right) ^2+\left( nh\right) ^2+\left(\frac{h}{2\pi}\theta\right)^2};\;\;\frac{\theta}{2\pi}\leq1.
    \end{equation}

   From the above formulas follows that when the effect of the stator  helicity on the
   self-inductance of the turn is strong, this effect should be checked against the difference between the self-inductance of
   the ring and the turn having a pitch $h$, which enables one to check  the approximation accuracy when the turns
   are replaced by the rings.

    \section{Resistance }

    \subsection{Skin Layer}
    {
    The skin-layer technique  is the most common method for calculating the HFCG resistance. This method is based on the assumption
    that the total current uniformly flows through the plate of
    thickness equal to the skin depth (see Fig.
    \ref{fig:resistence_skin_sloj}). The skin depth $\delta$ can
    be found by formula
         \begin{equation}
    \frac{\rho}{\delta}=k\sqrt{\frac{\frac{dI(t)}{dt}}{I(t)}},\label{eq:phormula_tolsini_skin_sloja}
    \end{equation}
    where  $\rho$ is the specific resistance ($1.72\cdot10^{-8}$ copper) and  $k=\sqrt{\mu_0\rho}$ is the coefficient, which for copper equals
    $1.47\cdot10^{-7}$.
    The resistance of the circuit is defined by the formula for the resistance of the ring:
    \begin{equation}
     R=\frac{2\pi r\rho (1+\alpha_TT)}{d\delta},
    \end{equation}
    where $r$ is the radius of the circuit, $d$ is the wire diameter (metal), $\alpha_T$ is the temperature coefficient
    (0.0043 for copper), and $T$ is the temperature  (it always equals zero when heating is ignored)

    \begin{figure}
         \centering
         \includegraphics[width=0.3\linewidth]{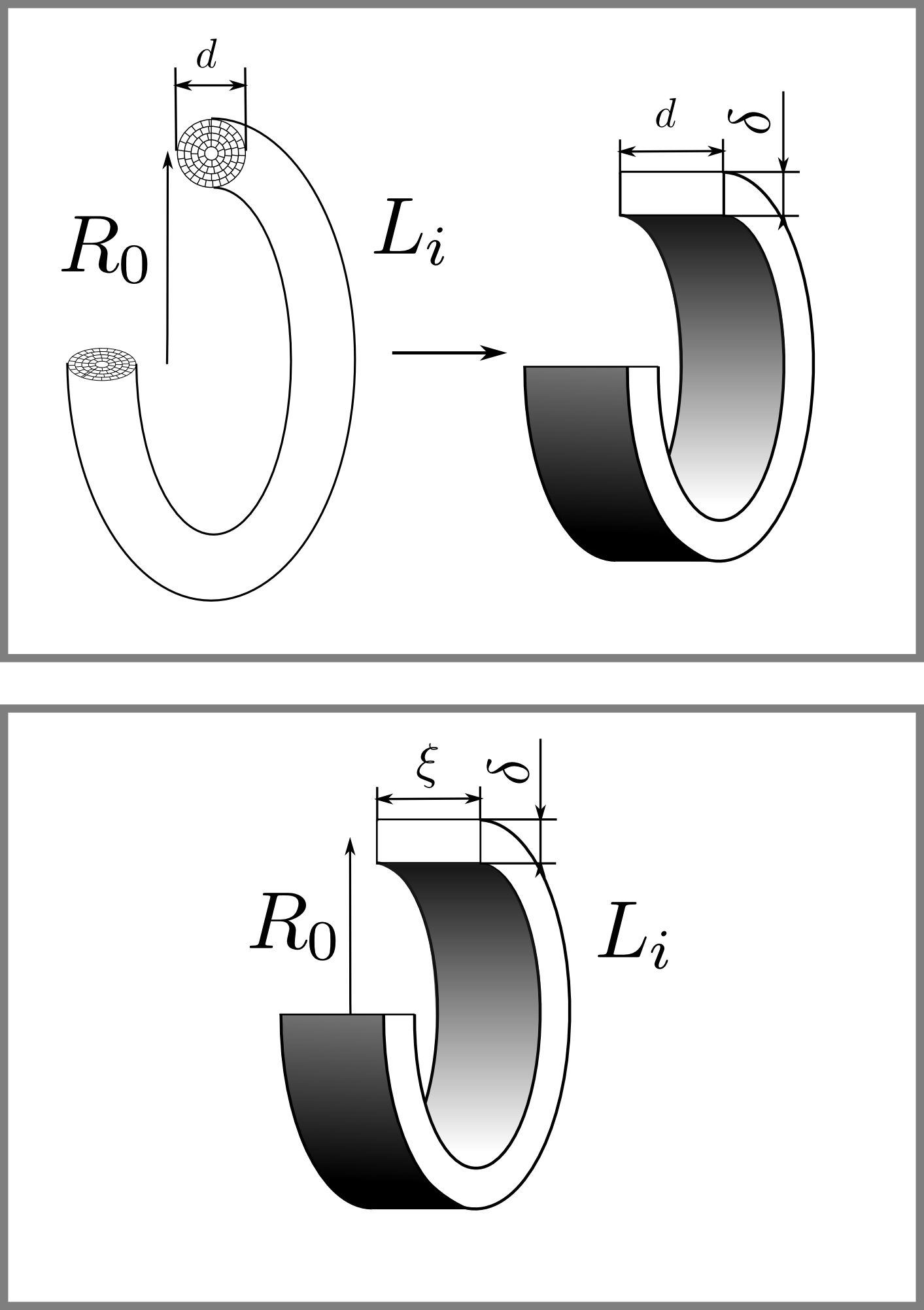}
                 \caption{The schematic representation of the resistance calculations using the skin-layer technique}
        \label{fig:resistence_skin_sloj}
    \end{figure}

    A similar resistance model for the circuits of the armature ring and the metal cup ring is defined by:
    \begin{equation}
     R= \frac{2\pi r\rho (1+\alpha_TT)}{\delta \cdot \xi },
    \end{equation}
    where $\xi$ is the width   of the armature ring and  the metal cup ring (in the general case, they can differ).

    To find the resistance $R_{line}$ of the armature and the metal cup to current
    $I_z$, let us  add together the corresponding resistances of
    the  armature rings and the metal cup rings to the current flowing in the
    direction of the OZ-axis:
         \begin{equation}
     R_{line}= \frac{\xi \cdot\rho(1+\alpha_T T)}{\pi(d\delta+\delta^2)},
    \end{equation}
    where $\pi(d\delta+\delta^2)$ corresponds to the area of the ring of radius $r$ and the width
    $\delta$.
    }

    \subsection{Nonlinear Magnetic Diffusion}

        In this subsection, we have  found the HFCG resistance from depth distribution of the current
    density in the stator turns and the armature rings.
    The current density distribution was determined from the
    equations for nonlinear magnetic diffusion into an infinite
    conducting plate ~\cite{Knopfel}:
            \begin{equation}
        \dfrac{\partial H_z}{\partial x}=-\dfrac{1}{\rho_0(1+\beta Q)}E_y=-j,\label{eq:rot_H}
        \end{equation}
        \begin{equation}
        \dfrac{\partial E_y}{\partial x}=-\mu_0 \dfrac{\partial H_z}{\partial t},\label{eq:diverative_H_t}
        \end{equation}
        \begin{equation}
        \dfrac{\partial Q}{\partial t}=(1+\beta Q)\rho_0 \left( \dfrac{\partial H_z}{\partial x}\right) ^{2}. \label{eq:specific_ohmic_losses}
        \end{equation}
   The initial conditions are as follows:

        $x=0: H_z(0,t)=H_0(t), t\geq0,$

        $t=0: H_z(x,0)=0, x\geq0.$

    Rewrite (\ref{eq:specific_ohmic_losses}) using (\ref{eq:rot_H})
        \begin{equation}
        \dfrac{\partial Q}{\partial t}=(1+\beta Q)\rho_0  j ^{2}
        \label{eq:specific_ohmic_losses_with_j},
        \end{equation}
    and  (\ref{eq:diverative_H_t}) using (\ref{eq:rot_H})
        \begin{equation}
        \dfrac{\partial (\rho_0 (1+\beta Q)j)}{\partial x}=-\mu_0 \dfrac{\partial H_z}{\partial
        t}\label{eq:diverative_H_t_with_j}.
        \end{equation}
    Let us write a more detailed expression for a partial
    derivative:
        \begin{equation}
        j \beta \rho_0 \dfrac{\partial Q}{\partial x}+ \rho_0 (1+\beta Q)\dfrac{\partial j}{\partial x}=-\mu_0 \dfrac{\partial H_z}{\partial t}.
        \end{equation}
    Here the initial and final conditions are as follows:
        $x=0: \dfrac{\partial H_z(0,t)}{\partial t} =f(t), t\geq0;$

        $t=0: H_z(x,0)=0,j(x,0)=0,Q(x,0)=0,  x\geq0.$

   Let us take  a flat plate of thickness $\frac{d}{2}$ and width $d$ (Fig.
   \ref{fig:fig1}), but we shall consider that the magnetic field
   is diffused into the infinite plate of finite thickness.

   In the absence of an external  magnetic field, the magnetic
   field strength near the plate surface is ~\cite{Kingsep}:
            \begin{equation}
        H=\dfrac{i}{2},
        \end{equation}
    where $i$ is the surface current density. The partial time derivative is
            \begin{equation}
        \dfrac{\partial H}{\partial t}=\dfrac {1}{2}\dfrac{\partial i}{\partial t}.
        \end{equation}
        Let $H_0$ denote the magnetic field strength in front of
        the plate in the case of an external magnetic field when
        a zero total magnetic field behind the plate
            \begin{equation}
        H_0=i.
        \end{equation}
 Take a time partial derivative
        \begin{equation}
        \dfrac{\partial H_0}{\partial t}=\dfrac{\partial i}{\partial t}.
        \end{equation}
    Divide the plate into $n$ number layers of thickness $h_x$ (Fig. \ref{fig:fig1}).
        \begin{figure}
             \centering
            \includegraphics[width=0.6\linewidth]{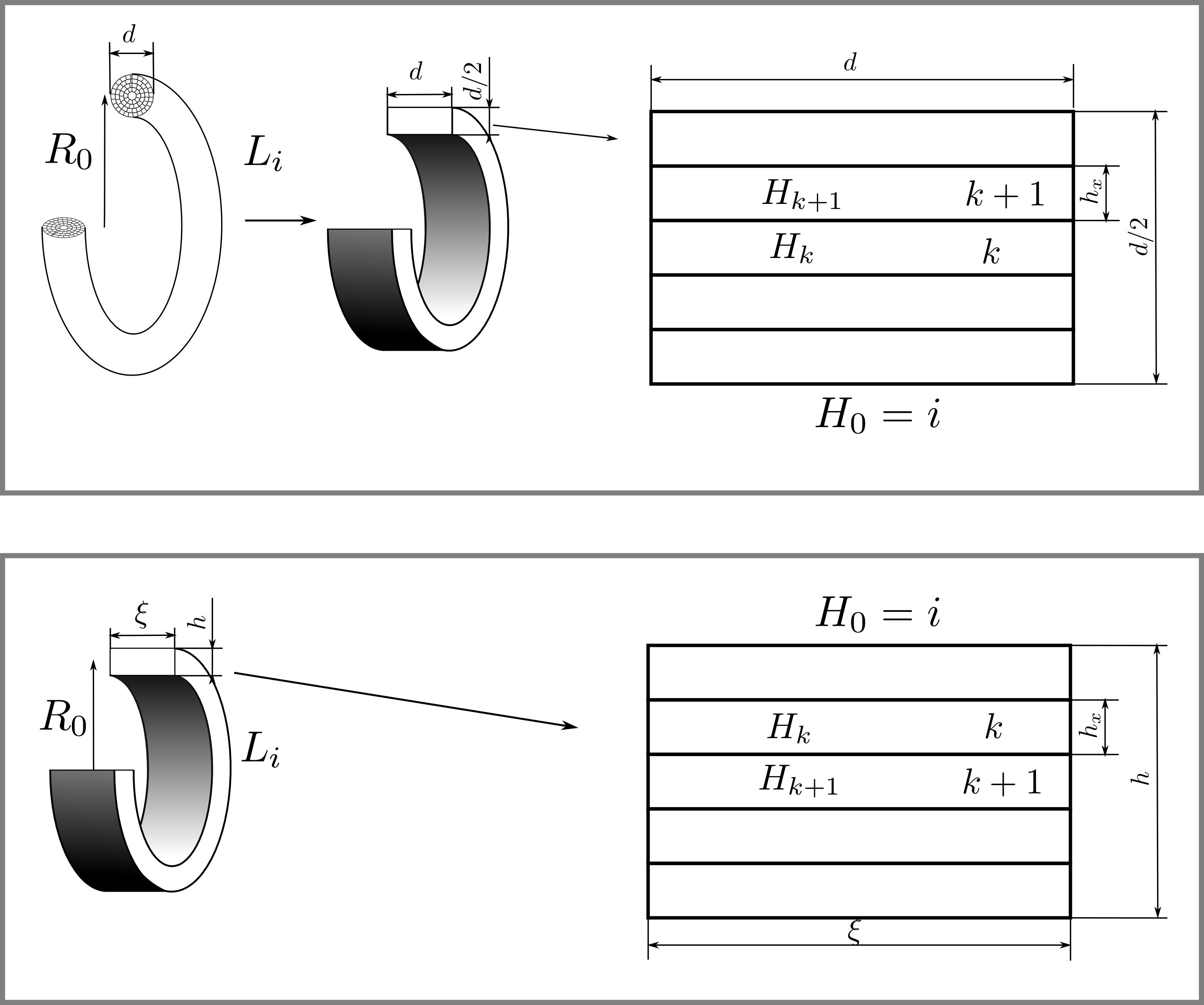}
                        \caption{Schematic representation of the plate division}
            \label{fig:fig1}
        \end{figure}

    The time  derivatives of the magnetic field strength in the adjacent layers of the plate are  related
    as follows:
            \begin{equation}
            \dfrac{\partial H_k}{\partial t}=\dfrac{\partial H_0}{\partial t}-\int\limits_0^{x_k}\dfrac{\partial j_k}{\partial t}dx\approx \dfrac{\partial H_0}{\partial t}-h_x\sum_{p=1}^{k}\dfrac{\partial j_p}{\partial t}.
        \end{equation}
    Define the boundary condition on the plate surface as
        \begin{equation}
            \dfrac{\partial H_0}{\partial t}=\dfrac{1}{d}\dfrac{\partial I}{\partial t},
        \end{equation}
    where $\dfrac{\partial I}{\partial t}$, the  current derivative in the equivalent
    circuit, is given in the computer code for HFCG calculations.
    The formula for the difference scheme can be written in the
    form:
            \begin{equation}
        \rho_0 \beta  j_{k+1}\dfrac{Q_{k+1} - Q_{k}}{h_x}+ \rho_0
        (1+\beta Q_{k+1})\dfrac{j_{k+1}-j_k}{h_x}=-\mu_0( \dfrac{\partial H_0}{\partial t}-h_x\sum_{p=1}^{k}\dfrac{\partial j_p}{\partial
        t}).
        \end{equation}

    From this one can find $\dfrac{\partial j_k}{\partial t}$  for $k=\overline{1..n-1}$
    and determine the nod values of current density  in the next time
    step. For $k=n$
        \begin{equation}
        \dfrac{\partial j_k}{\partial t}h_x=\dfrac{\partial H_0}{\partial t}-h_x\sum_{p=1}^{n-1}\dfrac{\partial j_p}{\partial t}.
        \end{equation}

        \begin{figure}
            \centering
            \includegraphics[width=0.3\linewidth]{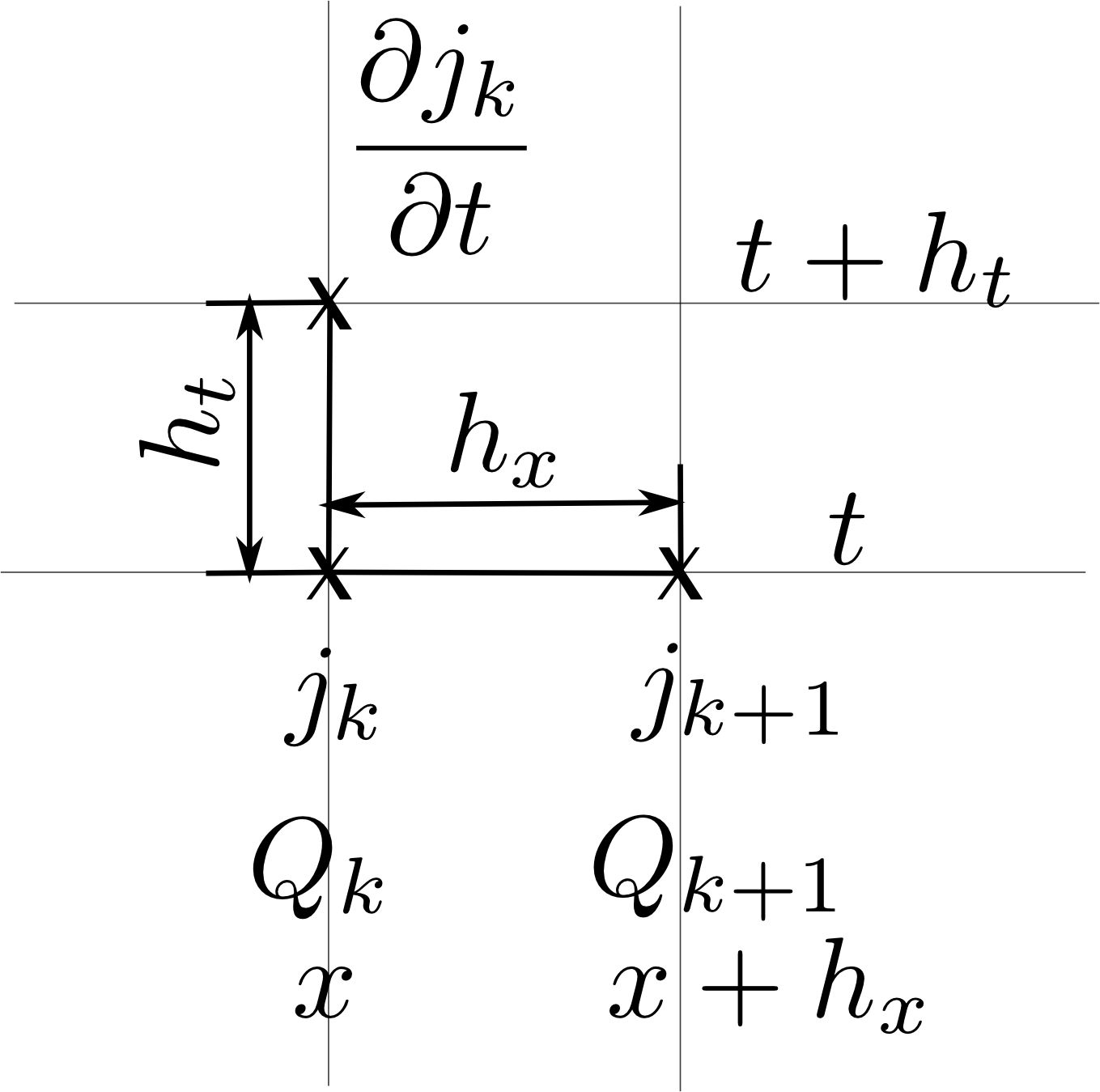}
                       \caption{Difference scheme}
            \label{fig:fig2}
        \end{figure}

  Since this difference scheme exhibits poor convergence and requires the use of very small integration steps,
  for the above formulas we shall write  the difference scheme  for the implicit Euler method:
    \begin{equation}
        \rho_0 \beta  j_{k+1}^{t+h_t}\dfrac{Q_{k+1}^t - Q_{k}^t}{h_x}+
        \rho_0 (1+\beta Q_{k+1}^t)\dfrac{j_{k+1}^{t+h_t}-j_k^{t+h_t}}{h_x}=-\mu_0( \dfrac{\partial H_0}{\partial t}
        -h_x\sum_{p=1}^{k}\dfrac{ j_{p}^{t+h_t}-j_{p}^t}{h_t}). \label{eq:raznostnaja_shema}
    \end{equation}
   Upon multiplying  by $\frac{h_t}{\mu_0h_x}$ and  transposing the terms containing the current
   density of the time step  $t+h_t$, the left-hand side of
   \eqref{eq:raznostnaja_shema} takes the form:
       \begin{equation}
        \frac{h_t}{\mu_0h_x}\rho_0 \beta  j_{k+1}^{t+h_t}\dfrac{Q_{k+1}^t - Q_{k}^t}{h_x}+ \frac{h_t}{\mu_0h_x}\rho_0 (1+\beta Q_{k+1}^t)\dfrac{j_{k+1}^{t+h_t}-j_k^{t+h_t}}{h_x}- \sum_{p=1}^{k} j_{p}^{t+h_t}.
    \end{equation}
        \begin{figure}
            \centering
            \includegraphics[width=0.3\linewidth]{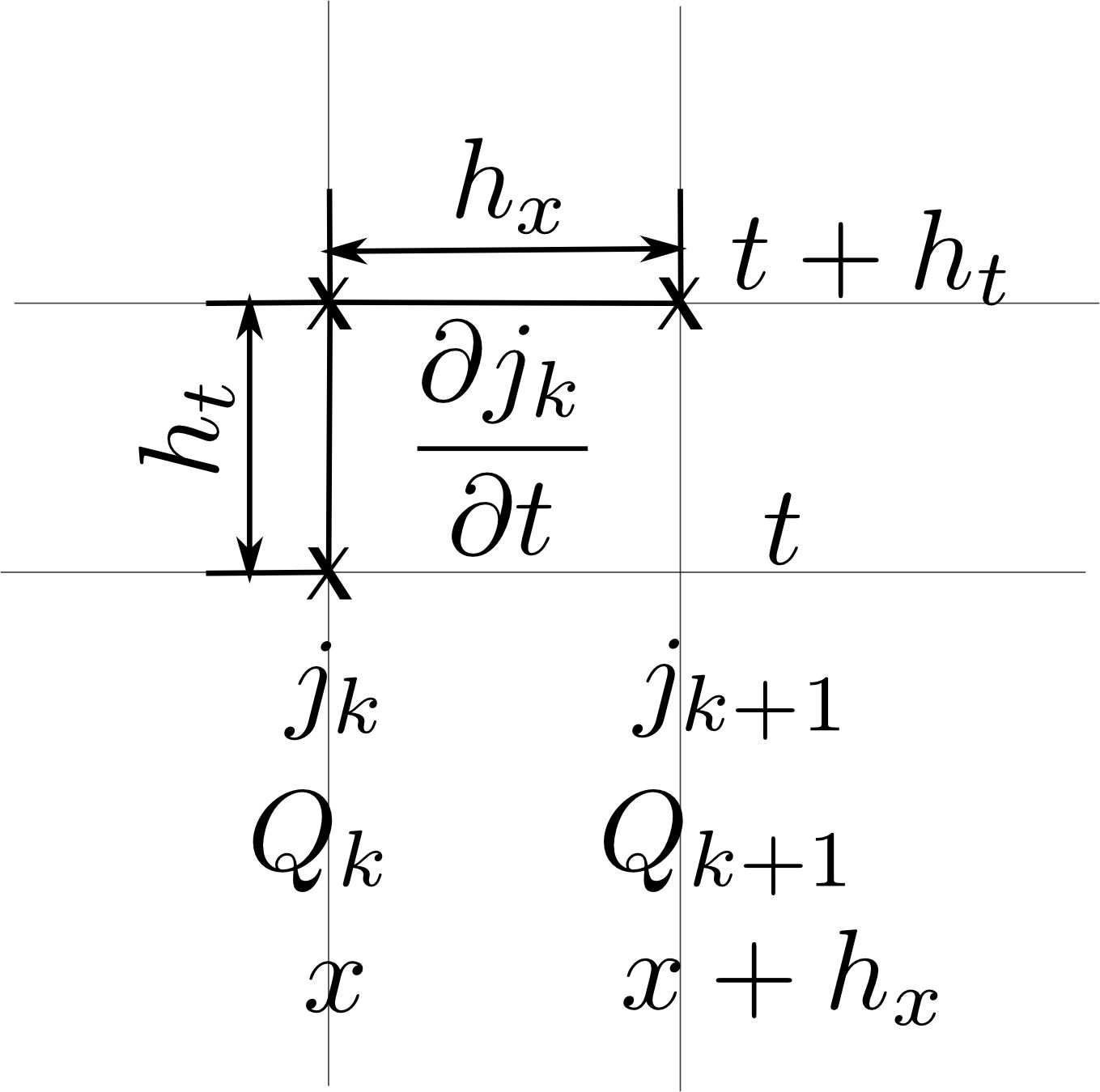}
                        \caption{The difference scheme for the implicit Euler method}
            \label{fig:razn_shema_nejavn_med}
        \end{figure}
    On the right-hand side of \eqref{eq:raznostnaja_shema}, we have:
    \begin{equation}
     -\frac{h_t}{h_x} \dfrac{\partial H_0}{\partial t}-\sum_{p=1}^{k} j_{p}^t.
    \end{equation}
 The difference scheme for the implicit Euler method is given in Fig. \ref{fig:razn_shema_nejavn_med}.
    Solving the obtained system of inhomogeneous linear equations,
    one can find the current density derivatives in each layer of
    the plate.
    Let us estimate the resistance. The formulas used in electric engineering for calculating resistance in parallel
    conductors  are not applicable because they require
    determining the resistance which occurs  in the plate layers
    due to the magnetic field change. The active resistance describes dissipative processes in the
    system, hence the resistance in the circuit can be determined
   with  known energy losses. Then for the general case, we can
   write:
            \begin{equation}
        Q^{t+h_t}_{total}-Q^{t}_{total}=RI^2h_t=\sum_{k=1}^n \left( Q_{k,total}^{t+h_t}-Q_{k,total}^{t}\right),
        \end{equation}
        where $Q^{t}_{total}$ is the electromagnetic field energy that has passed through the conductor
        during time $t$, $h_t$ is the integration time step, and $k$ is the subscript numbering
        the plates in the ring
           \begin{equation}
        R=\dfrac{\sum_{k=1}^n \left( Q_{k,total}^{t+h_t}-Q_{k,total}^{t}\right)}{I^2h_t}.
        \end{equation}

    In the integral representation of the energy equation \cite{Knopfel} two types of field
    losses are distinguished:
        \begin{equation}
     -\int\limits_S\textbf{P}ds=\int\limits_V\frac{\partial}{\partial
     t}(Q+W)dV,
    \end{equation}
    where $\frac{\partial Q}{\partial t}=\rho j^2$ is the Joule loss, $j$ is the current density,
     $\frac{\partial W}{\partial t}=\textbf{H}\frac{\partial \textbf{B}}{\partial t}=\frac{\partial }{\partial t}
     \left( \frac{1}{2}\mu \textbf{H}^2\right) $ is the change of the electromagnetic field energy, and
      $\textbf{P}=\left( \textbf{E}\times\textbf{H}\right) $ is the Poynting vector, i.e, the
      energy flux passing through the surface.

 Express the derivative of the magnetic field energy in terms
  of the surface current density:
       \begin{equation}
     \frac{\partial W}{\partial t}=\textbf{H}\frac{\partial \textbf{B}}{\partial t}=i\mu\frac{\partial i}{\partial t}.
    \end{equation}
    For each plate layer with the current, we can write
       \begin{equation}
     \frac{W_k^{t+h_t}-W_k^{t}}{h_t}=\textbf{H}_k\frac{\partial \textbf{B}_k}{\partial t}=i_k\mu\frac{\partial i_k}{\partial
     t}.
    \end{equation}
     Using \eqref{eq:specific_ohmic_losses_with_j}, let us find the energy released
     through Joule heating
        \begin{equation}
        Q_k^{t+h_t}=Q_k^{t}+\int\limits_t^{t+h_t}(1+\beta Q_k^{t})\rho_0 \left( j_k^{t}\right) ^{2}dt.
        \end{equation}
   Application of Simpson's rule for numerical integration  gives
        \begin{equation}
        Q_k^{t+h_t}=Q_k^{t}+(1+\beta Q_k^{t})\rho_0 h_t\left( \left( j_k^{t}\right) ^{2}+
        4\left( j_k^{t}+\dfrac{\partial j_k^{t+h_t}}{\partial t}    \dfrac{h_t}{2}\right) ^{2}+
        \left( j_k^{t}+\dfrac{\partial j_k^{t+h_t}}{\partial t} h_t\right) ^{2}\right)/6. \label{eq:specific_ohmic_losses_with_j_razn}
        \end{equation}
    Finally, we have for the resistance:
        \begin{equation}
        R=\dfrac{\sum\limits_{k=1}^n \left( Q_{k}^{t+h_t}-Q_{k}^{t}+W_k^{t+h_t}-W_k^{t}\right)}{I^2h_t}.\label{eq:efective_resistance_of_FCG}
        \end{equation}

    The obtained formula shows that the heating losses can be determined
    from the current density distribution. When the resistance is
    found from the skin depth, the losses related to the diffusion
    of the magnetic field into the conductor should be included
    along with the Joule losses. Thus, the accurate calculation of heating and temperature values seems to be more complicated.
     }
    \section{Voltage and Electric Field Strength in the HFCG}
    {
    Upon calculating $U_i$ from the system of equations (\ref{eq:uravnenie-s-razbivkoj}), one can find
    the voltage across each equivalent circuit of the stator  turn. As is
    seen from  the schematic diagram of the HFCG (Fig.
    \ref{fig:Electrical_diagram}), summation of the voltages across
    the turns with respect to the armature gives the
    stator-armature voltage over the HFCG length. The voltage
    produced by the current $I_z$ in the armature  can be neglected
    because it only amounts to several  per cent of the voltage
    across a single turn, being as small as the calculation error
    for the voltage across the turns in the vicinity of the
    contact point in a two-dimensional model.  With known voltage
    across the stator turns, one can make a map of the
    electric field strength for the stator-liner field upon
    calculating the distance between the wires of the turns. This
    distance is determined with regard for the size of the
    wire and the position of the liner at each instant of time
    Fig. \ref{fig:Electric_field_strength}.
             \begin{figure}
    \centering
        \includegraphics[width=0.6\linewidth]{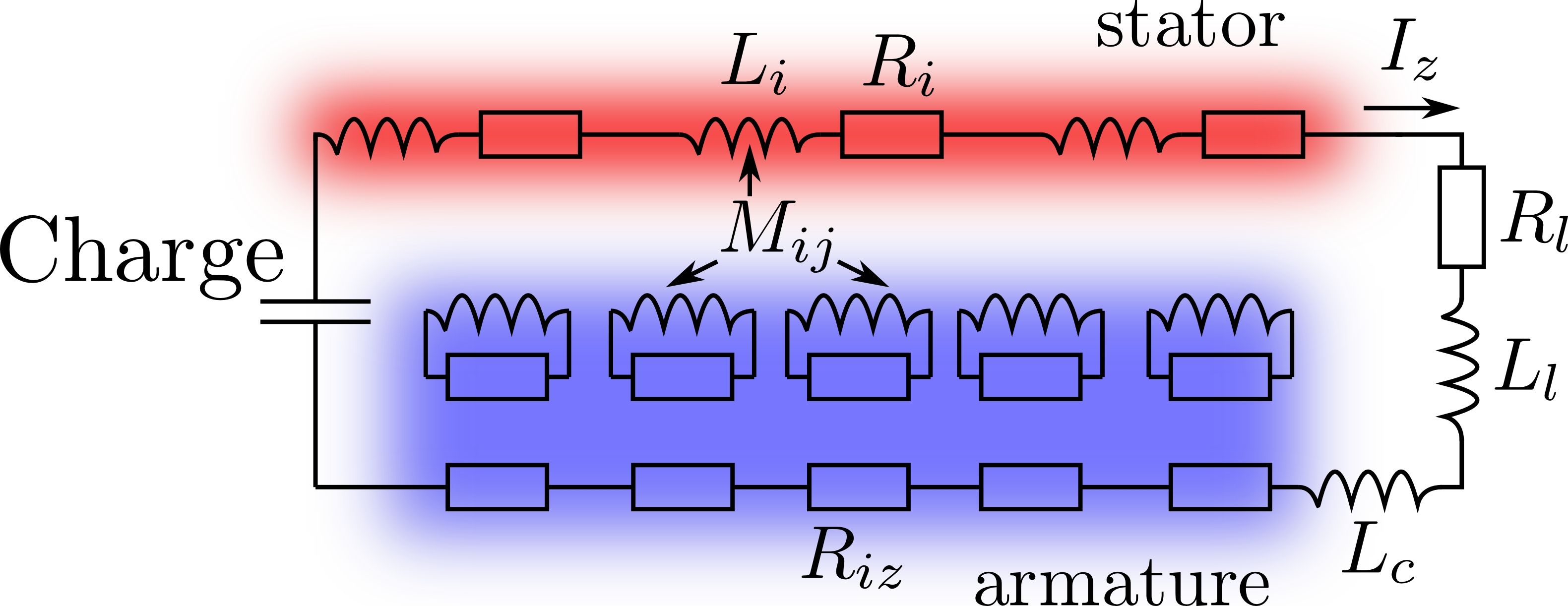}
               \caption{Electrotechnical scheme of the proposed 2D model}
        \label{fig:Electrical_diagram}
    \end{figure}

            \begin{figure}
    \centering
        \includegraphics[width=0.3\linewidth]{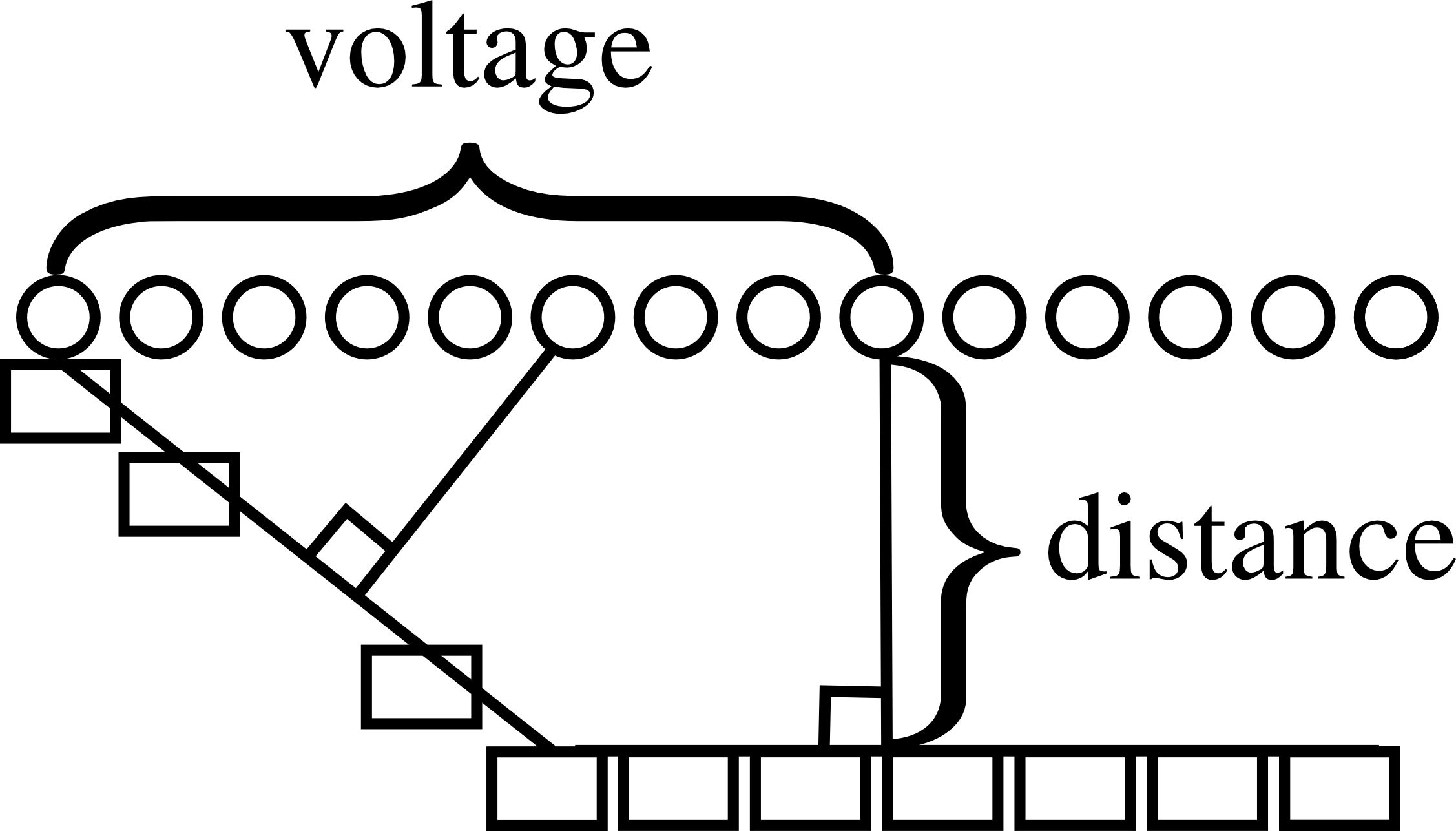}
                \caption{HFCG electric field strength calculation}
        \label{fig:Electric_field_strength}
    \end{figure}
    }
        \section{The Model of HFCG Seeding. Plastic and Metal Cups.}
    {Seed sources of different types are used for feeding the HFCG:
     capacitor banks, storage batteries, and other HFCGs. We shall
    consider the option where a capacitor bank is used as a seed
    source for its simplicity.
     The capacitor bank is charged
    first and then discharged into the HFCG via the connectors,
    thus causing the current gain in the HFCG. After high explosives
    are ignited, the stator or the metal cup is connected to the
    armature, the HFCG operates independently of the capacitor
    bank.

    Let us distinguish two HFCG designs according as a metal
    or plastic cup is used. For the case of a metal cup with the
    length comparable to the diameter of the stator, the armature
    gets connected to the metal cup far from the stator, and
    the derivative of the HFCG inductance is less than the losses
    in the circuit, so the current decreases.
    The current starts
    to increase when the vertex of the armature cone reaches the
    stator.

    From the viewpoint of electrical technology, the case when a plastic cup is
    used is similar to that without a metal cup. The stator is connected
    to the armature at the first turn. It occurs at the
    moment when feeding from the capacitor bank is stopped,
    and the bank continues feeding the HFCG until  the armature
    is connected to the stator. The  HFCG inductance
    changes during the seeding process, and the current derivative is
    usually positive (of the same sign). In this case, there is an
    option of whether to calculate the HFCG inductance from the
    magnetic flux or from the magnetic energy. These methods give different
    results for the derivative of the HFCG inductance, and the
    difference is most significant during the seeding process. For this reason, an accurate model of the process of seeding the
    HFCG with a plastic or a metal cup would enable one to choose
    a more appropriate method of the two, thus demonstrating its advantages.

    }

    \section{Conclusion}
    This paper describes a two-dimensional model of a helical FCG,
    based primarily on fundamental physical principles. The
    experimental results have been described for various HFCG
    designs and operation parameters with an accuracy within
    instrumental error.
    The idea stated in  \cite{Neuber} that
    the intrinsic losses in the magnetic flux are not of resistive character has
    been confirmed and the physical explanation for these losses
    has been suggested, as well as the method of their calculation.
    The model of nonlinear diffusion of the magnetic field into the conductor and
   the model of resistance have been developed.
   It has been shown
   that the  magnetic field diffusing into the conductor is the
   source of magnetic flux losses, which enhance the
   resistance of the conductor, and the resulting resistance  equals the resistance of the conductor calculated using the
   skin-depth technique.
   It has also been demonstrated that the
   method of elimination of the stator turns from the system of
   equations  \eqref{eq:Iz circuit}, \eqref{eq:theta circuit} and
   re-calculation of the current in the turn is redundant.
   The proposed model of the moving contact point
  enables one to  demonstrate the mechanism of the intrinsic flux losses and
   present a physically correct two-dimensional model of a helical
   FCG for the description and anticipation of experimental
   results.

However, some processes in a two-dimensional model are simplified,
thus limiting its application, for example, for the description of
an HFCG  whose stator coil has widely-spaced turns with
inter-turn spacing comparable to the diameter of the coil wire.
In this case, one should use a three-dimensional HFCG model, which
correctly accounts for the current density distribution over the
wire cross section.
It should be noted that the resistance of the stator turns can
differ as much as by a factor of three, depending on the distance
to the contact point, though the average resistance of the turns
is practically constant, which follows from the problems solved in
a 3D model \cite{We3D}. Development of a valid three-dimensional
model for HFCG description is the next step in theoretical
analysis of HFCGs.

    \newpage


\begin{thebibliography}{99}

        \bibitem{Sakharov}Sakharov A. D. et al. Sov. Phys. Dokl. 10 1045 (1966)
        \bibitem{Fortov}  Fortov V. E.   Explosive-driven Generators of Powerful Electric Current Pulses
 Cambridge International Science Publishing, 2003
        \bibitem{Kalantarov}Kalantarov P. L. and Zeitlin L. A. Calculation of Inductances: A Handbook,
Leningrad, Energoatomizdat, 1986 (in Russian).
        \bibitem{Knopfel} Knoepfel H. Pulsed High Magnetic Fields, North-Holland, Amsterdam 1977.
        \bibitem{Kingsep}Kingsep A. S.,  Lokshin G. R. and Ol'khov O. A.,Fundamentals of Physics.  Physics in 2 v. V. 1.
        Mechanics, Electricity and Magnetism, Waves, wave optika.uchebnik for universities.
 FIZMATLIT, 2001(in Russian).
        \bibitem{Novac}B.M. Novac, I.R. Smith, M.C. Enache, H.R. Stewardsom Simple 2D model for helical flux-compression generators, Laser and Particle Beams (1997), vol. 15 , no3, pp. 379-395
        \bibitem{Neuber} Neuber A. A. (editor) Explosively Driven Pulsed Power: Helical Magnetic Flux Compression Generators, Springer, Berlin, 2005. P. 280.
 \bibitem{Kiuttu} G. F. Kiuttu and J. B. Chase, An armature-stator contact resistance model for explosively driven
helical magnetic flux compression generators, Proc. 15th IEEE Intl. Pulsed Power Conf., Monterey
(2005).
 \bibitem{KiuttuLast}G.F. Kiuttu, J.B. Chase, D.M. Chato, G.G. Peterson Recent advances in modeling helical FCGs Proceedings of Megagauss XI, Santa Fe, Omnipress, pp. 255  264 (2008).
\bibitem{Crawford} J.C. Crawford and R.A Damerow, Explosively Driven High-Energy Generators, J. Appl. Phys., 39, No. 11, pp. 5224-5231 (1968).
\bibitem{Fowler} Fowler C.M., Garn W.B. and Caird R.S. Production of very high magnetic fields by implosion, J.Appl.Phys.  1960.  V. 31.  P. 588594.
\bibitem{We3D} Haurylavets V.V. and Tikhomirov V.V,  3-D modeling for losses in HFCGs, International student's forum
"First step in science - 2009", Minsk, Pravo i ekonomika, 2010,
v.2, p.630 (in Russian).  
\bibitem{Paul}Paul, Clayton R
Inductance: Loop and partial / by Clayton R Paul.-- New
Jersey: Wiley-IEEE Press, 2010. xiii, 379p.
ISBN : 9780470461884.
621.3742 P10 185367
        \end{thebibliography}
\end{document}